\documentclass[12pt,preprint]{aastex}
\begin{document}
\title{Kinematics of planetary nebulae in 
the outskirts of the elliptical galaxy NGC 4697\altaffilmark{1}}

\author{R. H. M\'endez\altaffilmark{2}, A. M. Teodorescu\altaffilmark{2}, 
R.-P. Kudritzki\altaffilmark{2} and A. Burkert\altaffilmark{3}}

\email{mendez@ifa.hawaii.edu}

\altaffiltext{1}{Based on data collected at the Subaru Telescope, which is 
operated by the National Astronomical Observatory of Japan.}

\altaffiltext{2}{Institute for Astronomy, University of Hawaii, 
                 2680 Woodlawn Drive, Honolulu, HI 96822}

\altaffiltext{3}{Universit\"ats-Sternwarte M\"unchen, Scheinerstr. 1, 
                 D-81679 M\"unchen, Germany}

\begin{abstract}
We describe the implementation of slitless radial velocity measurements 
of extragalactic planetary nebulae (PNs) with the 8.2 m Subaru telescope 
and its Cassegrain imaging spectrograph, FOCAS. As a first application, 
we have extended a previous search for PNs in NGC 4697 to larger angular 
distances from its center. A total of 218 PNs were detected, and their 
radial velocities were measured. We have added 56 new PN detections to
the existing sample of 535, observed previously with the ESO VLT + FORS 
imaging spectrograph; 36 of these new 56 PNs are located at angular 
distances larger than 230 arcsec from the center of NGC 4697. 
We compare the new FOCAS velocities with the earlier FORS 
velocities, for 158 of the 162 reobserved sources, finding 
good agreement. We now have kinematic information extending out to 
5 effective radii from the center. The outer line-of-sight velocity 
dispersion is a bit lower than estimated earlier. This result is
compatible with the existence of a dark matter halo plus some degree
of radial anisotropy, but the dark matter halo is rather inconspicuous,
and it is still unclear how massive it can be.
A more detailed global dynamical study of the whole set of PN velocities 
will be required to decide if they permit to narrow down the range of 
possible dark matter distributions in NGC 4697. The new radial velocities 
reveal no evidence of rotation at 5 effective radii. 
\end{abstract}

\keywords{galaxies: elliptical and lenticular, cD --- 
galaxies: individual (NGC 4697) --- galaxies: kinematics and dynamics ---
planetary nebulae: general --- techniques: radial velocities}

\section{Introduction}

Planetary nebulae (PNs) in the outskirts of elliptical galaxies 
offer valuable kinematic information that can be used to trace the 
angular momentum distribution and mass distribution in these 
galaxies. The classic example is the study of PNs in NGC 5128 
by Hui et al. (1995), later extended by Peng et al. (2004) to 
780 PNs. A dark matter halo and significant outer rotation were 
detected. A recent dynamical study of 340 globular clusters in 
NGC 5128 has shown additional evidence of a dark matter halo 
(Woodley et al. 2007). Another case where a dark matter halo was 
immediately apparent from PN kinematics is NGC 1344 (Teodorescu 
et al. 2005).

In a previous work (M\'endez et al. 2001), hereafter Paper I, 
we presented photometry and slitless radial velocity 
information for 535 PNs in the flattened elliptical galaxy 
NGC 4697. These PNs provided kinematic information up to a distance 
of 3 effective radii from the center of NGC 4697. The most important 
result in Paper I was a substantial decline of 
line-of-sight velocity dispersion (losvd) as a function of angular 
distance from the center, which can be interpreted as indicating 
the absence of a dark matter halo around this galaxy,
if we assume an isotropic velocity distribution. 
However, some dark matter can be present if the velocity 
distribution is radially anisotropic; we explicitly warned about 
this alternative interpretation in Paper I, and the point 
has been made in much more detail by Dekel et al. (2005) and
De Lorenzi et al. (2008a).

It is important to clarify which of the two interpretations,
lack of dark matter or pronounced radial anisotropy, is correct. In 
the current context, with dark matter firmly established as one of 
the basic components of the universe, the existence of one or several 
substantial, medium-size elliptical galaxies without a dark matter 
halo (which is expected to have made possible its very formation) 
is surprising. On the other hand, if we are able to confirm 
radial anisotropy, then we have 
learned about an important constraint concerning the formation of 
elliptical galaxies. For example, Athanassoula (2005) has made 
numerical experiments in which a quick succession of disk mergers 
is more likely to produce significant radial anisotropy.

De Lorenzi et al. (2008a) made a detailed dynamical analysis
of NGC 4697, using the PNs of Paper I and long-slit absorption line 
kinematic data. Their best-fitting models were radially anisotropic, 
and included potentials with a variety of dark matter halos. The models
with no dark matter were not consistent with the PN losvds. However,
more PN velocities at larger angular distances from the galaxy's 
center were required to narrow down the range of acceptable dark 
matter models. In this paper we provide more such PN velocities. 

Romanowsky et al. (2003) showed similarly declining losvds in 
NGC 3379, NGC 4494 and NGC 821, claiming that their ``orbit library 
analysis'' ruled out extreme orbital anisotropies at least in the 
case of NGC 3379. More recent modeling (De Lorenzi et al. 2008b) 
indicates again that there 
are mass-shape-anisotropy degeneracies that need to be resolved to 
reach a firm conclusion. In addition, a study of two dozens of 
globular clusters in NGC 3379 (Pierce et al. 2006) suggests 
that some dark matter may be present there.  
A recent long-slit spectroscopic study by Forestell \& Gebhardt 
(2008) seems to indicate the presence of a dark matter halo 
around NGC 821, again in contradiction to what Romanowsky et al. 
(2003) concluded from the PNs in that galaxy. In 
summary, the case for no dark matter appears to be weakening,
but it is still quite possible that less massive ellipticals have, 
on average, lower dark matter fractions, i.e. less easily detected 
dark matter halos (e.g. Napolitano et al. 2005). We seem to be
in a situation that will require more careful studies of several 
galaxies before we can arrive to more solid conclusions.

The need to collect large PN samples means that we need to detect
faint PNs. This can be better done using the largest telescopes 
available; so we had a strong motivation to extend the slitless
radial velocity method, used in Paper I with VLT+FORS, to other 
large telescopes in the northern hemisphere. The FOCAS spectrograph 
of Subaru proved to be a good choice. As a first application of the
slitless technique with FOCAS and Subaru, we have extended our 
search for PNs around NGC 4697 to larger angular distances. In this
paper we present the resulting data and a preliminary analysis. 
In Section 2 we review the basic idea of slitless radial velocities
and describe the Subaru + FOCAS observations in the outskirts of 
NGC 4697. Section 3 describes the reduction and calibration procedures.
Section 4 deals with quality control of the slitless radial velocities 
using observations of the local PN NGC 7293. In Section 5 we give a
catalog of 218 PNs detected with FOCAS, and compare the FOCAS with 
FORS slitless radial velocities for 158 PNs. Having verified the 
quality of the data, in Section 6 we discuss the line-of-sight
velocity dispersion, and we analyze a plot of the escape 
velocity as a function of projected distance from the center of the 
galaxy. We adopt a distance of 10.5 Mpc for NGC 4697, taken from 
Paper I. Section 7 deals with rotation in the outskirts of NGC 4697. 
Section 8 gives a summary of results.

\section{Radial velocity method, field selection and observations}

Since the slitless radial velocity method was described in 
Paper I, here we give only a brief summary of the basic ideas.
Planetary nebulae can be detected in the light of [O III] $\lambda$5007 
using the traditional on-band, off-band filter technique. Having taken 
the on-band image, insertion of a grism as dispersing element produces 
not only dispersion but also a displacement of all images. 
The dispersed images of PNs remain point sources, 
which permits an accurate measurement of the displacement. Calibration 
of the displacement as a function of wavelength and position in the CCD 
detector offers an efficient way of measuring radial velocities for all 
PNs in the field, irrespective of their number and distribution. 
Since no slits are used, there are no light losses, and it is not 
necessary to go through the complex selection and preparation 
procedures typical of multi-object slit (or fiber) spectroscopy. 
Of course the quality of the slitless velocities will depend on the 
seeing and will decrease for very faint sources because of the larger
position uncertainty. 

The project reported here involves the 8.2 m Subaru telescope on Mauna 
Kea, Hawaii, and its Cassegrain imaging spectrograph FOCAS. We wanted 
to extend our search for PNs to larger angular distances from the 
center of NGC 4697, but allowing for partial overlap with the
FORS E and W fields described in Paper I, to be able to compare
the radial velocities measured with both instruments. Figure 1
shows the FORS fields studied in Paper I, and the two FOCAS 
fields observed for the present work.

The Subaru + FOCAS observations were made on the nights of
2004 May 12/13, 13/14, and 2005 May 11/12, 12/13 and 13/14.
The 2004 nights were partly clouded. The first two 2005 nights were 
photometric, and the last one was affected by thin cirrus. During
both observing runs the seeing was stable, oscillating around 0.6 
arcsec.

FOCAS has been described by Kashikawa et al. (2002). The circular
field of view of FOCAS has a diameter of 6 arc minutes. This field 
is covered by two CCDs of 2k$\times$4k (pixel size 15$\mu$m) with
an unexposed gap of 5 arcsec between them. The image scale is 
0.1 arcsec pix$^{-1}$. Two fields, E and W of NGC 4697, were 
observed, as shown in Figure 1.

On-band direct imaging was done through the 
interference filter N502, which has the following characteristics:
central wavelength 5025 \AA, peak transmission 0.68, and FWHM 60 \AA.
Off-band imaging was done through the broad-band standard visual filter.
Dispersed images were obtained by exposing through filter N502 and an 
echelle grism with 175 grooves mm$^{-1}$, operating in the 4th order.
In this way the echelle grism gives a dispersion of 0.5 \AA \ pix$^{-1}$, 
with good efficiency and a rather small displacement. Dispersed PN 
images of 0.5 arcsec have a size of 5 pixels on the CCD. This implies 
a radial velocity resolution of 140 km s$^{-1}$, i.e. the PN's
internal velocity field is not resolved. In other words, PNs should 
always appear as single point sources. Assuming position errors of 0.4 
pixel, the expected uncertainty in radial velocity is 10 km s$^{-1}$.

Table 1 lists the most important CCD images obtained for this project.
It includes the exposures of the FOCAS E and W NGC 4697 fields, the
spectrophotometric standard G138-31 (Oke 1990), and 
the calibration and quality control exposures to be described in next 
Section. Since there are two CCD images for each field (Chips 1 and 2), 
for brevity we have listed only the exposure number corresponding to 
Chip 1. For Chip 2, add 1 to the listed exposure number. All images 
were binned 2$\times$1 in the horizontal direction (perpendicular to 
dispersion). This was done to increase the signal from
the very faint sources we wanted to detect, but at the same time 
without compromising the spectral resolution. As a consequence of 
this binning, each point source looks like an ellipse. Unbinning 
is required to bring all stars back to their normal appearance.

\section{Reduction and calibration procedures}

The basic CCD reductions (bias subtraction, flat-field correction
using twilight flats) were made using IRAF\footnote{IRAF is
distributed by the National Optical Astronomical Observatories,
operated by the Association of Universities for Research in
Astronomy, Inc., under contract to the National Science
Foundation} standard tasks. Then we unbinned the images, recovering 
the original CCD pixel numbers in the horizontal direction. 

For image registration and combination we first selected, for each of the
E and W fields, a pair (undispersed, dispersed) of on-band individual 
images, of the best possible quality, taken consecutively at the 
telescope, and adopted them as reference images. For field E, chip 1, 
the reference images were 60989 and 60991. For field W, chip 1, the 
reference images were 52471 and 52473. Obviously, for chip 2 the 
reference images were 60990, 60992, 52472 and 52474. All the other 
images, including the off-band ones, were registered upon the 
corresponding reference image. Registration and combination were
made using IRAF standard tasks. In particular the dispersed images
were registered using the brightest PN images as reference point 
sources, avoiding in this way any dependence on ambient temperature 
variations (see the corresponding discussion in Section 3 of Paper I).

We still have to describe the calibration images for radial velocities. 
They were obtained using an engineering mask with punched holes. The 
mask produces an array of point sources covering the whole field 
when illuminated by the internal FOCAS lamps or by any extended 
astronomical source. There are almost 1000 calibration points,
separated by about 100 pixels in $x$ and $y$. Two kinds of calibration 
images were obtained illuminating the engineering mask with a Th-Ar 
comparison lamp: (u1) undispersed, taken through filter N502; 
(d1) dispersed, taken through N502 and the echelle grism. 
In addition, there were ``quality control'' images, obtained 
illuminating the engineering mask with the local PN NGC 7293:
(u2) undispersed, taken through N502; (d2) dispersed, taken through 
N502 and the echelle grism. Figure 2 shows examples of images u2,
d2 and d1. 

A comparison between the left and central images in Figure 2 shows 
the effect of inserting the grism; comparing to the position of a 
fixed CCD defect, easily recognizable, it is clear that all the point 
sources have been displaced upwards. The right image in Figure 2 shows 
what happens if we keep the grism in the light path and illuminate the 
mask with the Thorium-Argon comparison lamp; now each point source gives 
a Th-Ar spectrum. The positions of the Th emission lines permit to obtain
the wavelength calibration for each grid point. The Th-Ar lamp is ideally 
suited for our purposes, because its spectrum shows several emission lines 
within the on-band filter transmission curve. In Paper I we needed to 
take the calibration images through a broader filter, because the FORS 
spectrographs do not have Th-Ar lamps.

Measurement of the undispersed and dispersed calibration images permits 
to obtain the displacement produced by the insertion of the grism as a 
function of wavelength and position on the CCD. The slitless radial 
velocity calculation proceeds in the following way: (1) identify the 
four calibration points closest to the undispersed position of the PN;
(2) from the displacement between undispersed and dispersed PN images,
calculate the redshifted wavelength of the [O III] emission at each
of the four calibration points; (3) obtain the final wavelength by 
bilinear interpolation, weighting the four wavelength values according 
to the distance from each of the four calibration points to the PN;
(4) calculate the radial velocity from the final wavelength and the
heliocentric correction.

\section{Radial velocity quality control using NGC 7293}

The NGC 7293 images we obtained permit to measure radial velocities 
in two ways: (a) slitless, using images u2 and d2 to obtain the 
displacement, and using the displacement as a measure of wavelength 
and therefore velocity, in the way we described at the end of the 
previous section; (b) ``classical'', using images d1 and d2, and 
treating each mask hole as a slit. Figure 3 shows a comparison of 
slitless vs. classical radial velocities of NGC 7293. There is good 
agreement, which indicates that the calibration of the displacement 
works very well. In addition, the average heliocentric velocity of 
NGC 7293 from all the grid points turns out to be approximately $-$30 
km s$^{-1}$, in good agreement with the known systemic velocity
($-$27 km s$^{-1}$, measured by Meaburn et al. 2005). Furthermore,
individual grid points show a range of velocities, from $-$15 to $-$50
km s$^{-1}$. Clearly in some cases the mask holes have isolated gas 
parcels that are moving predominantly towards or away from us, while
in most cases we see gas moving in both directions, so that the 
unresolved emission line profile gives us a velocity closer to the 
systemic velocity. From the observed range we infer a lower limit of 
about 20 km s$^{-1}$ for the expansion velocity in [O III], again 
perfectly consistent with the 25 km s$^{-1}$ measured by Meaburn 
et al. (2005). No grid point has given a radial velocity departing
more from the systemic velocity than we should expect from the expansion 
velocity. From all these results we estimate that the uncertainty
in FOCAS slitless velocities is of the order of 10 kms$^{-1}$. In
the next section we will provide more evidence about the quality of
the FOCAS slitless velocities.

\section{PN detection, photometry, catalog, and comparison with FORS 
photometry and radial velocities}

We searched for PNs using the traditional blinking method.
In the same way as in Paper I, a PN candidate was found by blinking 
on-band vs. off-band, and confirmed by blinking undispersed vs. 
dispersed on-band. The source was accepted as a PN candidate if it 
was absent in the off-band image, and if it was a point source.
A total of 218 PNs were found, 162 of which had been discovered 
previously with FORS. Consequently, we have added 56 new 
detections. Most of the new detections (36 in fact) are at angular 
distances larger than 230 arcsec from the center of NGC 4697.
 
After measuring the positions of the 218 PNs on
the undispersed and dispersed images, the radial velocities were 
calculated as described in sections 2 and 3. 
We also measured the Jacoby magnitudes $m$(5007) using the same
spectrophotometric standard (G 138-31, Oke 1990) and the same 
procedures described in Paper I. For a definition of Jacoby 
magnitudes, see Jacoby (1989) or Paper I. The standard G 138-31 
has $m$(5007)=19.36 (this number depends on the characteristics 
of the on-band filter). The faintest objects detected both 
in Paper I and in the present search have $m(5007)\approx 28.5$.
We expect both PN searches to have similar limiting magnitudes, 
because the telescope sizes, instrumental and CCD efficiencies, 
and total exposure times, are similar.

J2000 equatorial coordinates for all PNs were calculated using a set 
of astrometric programs written by David Tholen and kindly provided 
by Fabrizio Bernardi. A first program, when given an input file 
with field and camera parameters, identifies all the USNO-B1
catalog stars available within the desired field and produces a list
of reference stars with rough estimates of their position in the chip.
The next programs make an improved centroid fitting for reference 
stars, producing an output file with a list of reference stars 
with pixel coordinates $x, y$ \/ accurate to a few hundredths
of a pixel. The last program performs the final astrometric fit,
rejecting outliers and iterating, and provides the PN equatorial 
coordinates with uncertainties of about 0.5 arcsec. 

In Table 2 we list the following for all detected PNs: identification 
number; ($x$, $y$) coordinates, in arcsec, relative to the optical
center of NGC 4697; J2000 equatorial coordinates; Jacoby magnitude,
$m$(5007); heliocentric radial velocity in km s$^{-1}$; 
and E and W FORS field identification numbers, if the PN is present 
in the FORS PN catalog (M\'endez et al. 2008). FOCAS identification 
numbers that begin with 20, 21, 22 and 23 correspond respectively 
to FOCAS fields W chip 1, W chip 2, E chip 1, and E chip 2. A few 
numbers are missing for a variety of reasons: object 2008's radial 
velocity is 281 km s$^{-1}$, if the emission line is identified as 
[O III] $\lambda$5007. Since this velocity is incompatible with 
NGC 4697, we assume that object 2008 is a background emission-line 
galaxy, at such a redshift that some other emission line falls into 
the on-band filter transmission curve (e.g. Ly$\alpha$ at $z$=3.1).
Object 2023 was a rediscovery of 2014. Object 2049 was not 
convincingly identified in the grism image, and was rejected as a PN.

Since we purposely reobserved many PNs discovered previously with 
FORS, we can test the accuracy of equatorial coordinates, Jacoby 
magnitudes, and, most important,
radial velocities. Figure 4 shows the differences in declination as a 
function of the differences in right ascension. We conclude that our 
coordinates are reliable to within one arcsec, which is adequate for 
our purposes.

Figure 5 shows the differences in magnitudes $m$(5007) plotted as a
function of the FORS $m$(5007), which is taken from M\'endez et al. 
(2008).
There is no systematic difference, and the dispersion increases, as 
expected, towards fainter magnitudes. We find no need to rediscuss
the photometric results and PNLF distance of 10.5 Mpc obtained in 
Paper I.

Figure 6 (left panel) shows FORS radial velocities as a function of 
FOCAS radial velocities for 158 PNs. There is satisfactory agreement. 
We should explain that four of the 162 PNs could not be measured in 
Paper I, because these PNs were too far to the right of the W FORS 
field (see Section 6 in Paper I).

We have generated simulations of radial velocity sets affected by random 
errors that follow a gaussian distribution. In Paper I we estimated
errors of about 35 or 40 km s$^{-1}$ for the FORS velocities. The right
panel in Figure 6 shows what happens if we assume FOCAS errors of the 
same magnitude. The dispersion is too large. The central panel in Figure
6 was obtained assuming FOCAS errors of 20 km s$^{-1}$. This is a much
better representation of what we observe in the left panel, and we 
conclude that The FOCAS errors are not larger than 20 km s$^{-1}$;
in other words, the FOCAS errors are significantly smaller than the 
FORS errors. This result is expected.
The FOCAS radial velocities are better mainly because the spectral 
dispersion is larger: 0.5 \AA \ per pixel versus the FORS dispersion 
of 1.2 \AA \ per pixel. For similar position measurement errors, 
the larger dispersion gives more accurate velocities.
Figure 7 shows the radial velocity difference FOCAS $-$ FORS
as a function of the apparent Jacoby magnitude $m$(5007) 
for 158 PNs. There is a small systematic 
difference of about 15 km s$^{-1}$, in the sense that FORS velocities
are larger, but it is well within the uncertainty attributed to FORS 
velocities. Note also that large differences occur only for faint PNs.
This is expected, because the quality of the slitless radial velocities
depends on the quality of the position measurements. For very faint 
sources, the positions in undispersed and dispersed images are less 
reliable because the sources are barely detectable and the S/N ratio 
is low.

\section{Line-of-sight velocity dispersions and escape velocities}

As mentioned in the introduction, the detailed dynamical analysis
by De Lorenzi et al. (2008a) showed that the best-fitting models 
are radially anisotropic, and require some dark matter; but it is
not clear how much. What we need is more PN radial velocities at 
large angular distances from the center of NGC 4697. Here we 
present a preliminary discussion of the new FOCAS PN velocities.
To illustrate how much we have improved the information about 
PNs distant from the center, we can mention that 
we have discovered a total (FORS + FOCAS) of 42 PNs at angular 
distances larger than 280 arcsec from the center of NGC 4697.
We have FORS velocities for 10 of these outlying PNs, and we have
FOCAS velocities for 41 of them. 
Figure 8 shows the positions of all FORS and FOCAS PNs relative
to the center of light of NGC 4697 (left panel), and their radial 
velocities as a function of the $x$ coordinates (right panel).
Our coordinate system is defined with the $x$-axis
coincident with the major axis of NGC 4697, and with the origin at 
the center of the galaxy, measuring $x$ and $y$ in arcsec.
Since the FORS data were discussed by De Lorenzi et al. (2008a), 
here we concentrate mostly on the new FOCAS velocities. 

In a first attempt we defined six data 
groups: PNs with $x < -300$, $-300 < x < -200$,  $-200 < x < -100$,
$100 < x < 200$,  $200 < x < 300$, and $300 < x$. The numbers of PNs 
within each group were, respectively, 14, 28, 49, 47, 30 and 8.
We calculated the line-of-sight velocity dispersions (losvds), and 
corrected them for the small effect of measurement errors by subtracting 
10 km s$^{-1}$ in quadrature. Figure 9 shows the resulting losvd for 
each group, as a function of average angular distance from the center 
of NGC 4697. We also show dispersions derived from absorption-line 
spectra along the major axis by Binney et al. (1990), and 
dispersions derived from FORS PNs in Paper I. All these data are in 
good agreement with each other where they overlap. 

De Lorenzi et al. (2008a) defined zones with elliptical boundaries for 
losvd calculations; see their Figure 6. Their outermost datapoint, at 
a radius of 300 arcsec, has a losvd of 110 km s$^{-1}$.
We have calculated the losvd for a group involving all the 48 
FOCAS PNs that lie outside an ellipse with semi-major and minor 
axes of 280 and 180 arcsec, respectively. The resulting losvd is 
93$\pm$13 km s$^{-1}$, somewhat smaller than the De Lorenzi losvd.
We made a similar calculation for the 21 FOCAS PNs lying outside a 
larger ellipse, with semi-major and minor axes of 350 and 225 arcsec.
The corresponding losvd is even smaller: 81$\pm$18 km s$^{-1}$.
These values are similar to those shown in Figure 9. We conclude, 
first, that our FOCAS losvds in the outskirts of NGC 4697 are not very 
sensitive to the shape of the regions used to define the samples; and
second, that the new FOCAS data indicate a slightly smaller losvd 
beyond 250 arcsec than obtained from the FORS data in De Lorenzi 
et al. (2008a). Following their discussion, we conclude that the best 
fitting models are still those including a not very massive dark matter 
halo, but it is still difficult to narrow down the range of acceptable
dark matter models, because the new FOCAS data lie somewhat closer to 
the models without dark matter than the earlier FORS data.

We also built a new sample by adding, to the 21 FOCAS PNs mentioned 
above, the six other FORS PNs that lie outside of the larger ellipse. 
We did not make any correction for the larger errors of those few 
FORS velocities. Still, 
the losvd for this new 27-object sample was 95$\pm$18 km s$^{-1}$, 
again lower than the De Lorenzi et al. value.

We can also consider an argument 
used in the pioneering work of Hui et al. (1995) to 
confirm the existence of a dark matter halo around NGC 5128. When we
plot PN radial velocities as a function of angular distance from the 
center of that galaxy, we find several PNs with velocities exceeding
the local escape velocity for a Hernquist (1990) model with constant 
$M/L$ ratio; see their Figure 20. The argument was repeated and 
improved in Peng et al. (2004); see their Figure 12. Using a constant 
$M/L$ Hernquist model, several PNs in NGC 5128 are unbound; we would 
expect such stars to escape in a very short time. Instead, using a 
two-component Hernquist model, which includes a dark matter halo,
even the most distant PNs become bound, because the escape velocity 
becomes much larger. This is a test we can easily apply to NGC 4697.
Figure 10 shows the result. We calculated the escape velocity for a 
constant $M/L$ Hernquist (1990) model at a distance $r$ from the 
center of the mass distribution:

$$   V_{\rm esc} = (2 G M_{\rm t} / (r + a))^{0.5},     \eqno(1)   $$

\noindent where $M_{\rm t}$ is the total mass (for our test we use two 
masses, 1.5$\times 10^{11} M_\odot$ and 9$\times 10^{10} M_\odot$), 
and $a$ is a scale length equal to $R_{\rm e}/1.8153$. 

The choice of galaxy mass needs to be explained. De Lorenzi et al. 
(2008a) found a $M/L$ ratio of about 5 in the R band, which is
equivalent to approximately 7.5 in the B band (see e.g. Maraston 1998).
Knowing the extinction-corrected $B_T$=10.0, the distance modulus 30.1
from Paper I, and the solar $B$ absolute magnitude 5.48, we obtain 
for NGC 4697 a blue luminosity of $1.7 \times 10^{10} L_\odot$, which
then gives a mass of $1.3 \times 10^{11} M_\odot$.

Note that the $(M/L)_B$ ratio of 11 determined in Paper I needs to be 
changed, because we have reduced the effective radius $R_{\rm e}$ of 
NGC 4697 after the photometric study by De Lorenzi et al. (2008a). 
The new value of $R_{\rm e}$ is 66 arcsec, which is equivalent to 
3.36 kpc at the galaxy's distance of 10.5 Mpc. This forces a decrease 
in the total mass required for a fit of the losvd with the Hernquist 
model. Please refer to formulas (32) and (41) in Hernquist (1990). 
The dispersion depends on the mass and the scale length $a$. If we 
change $R_{\rm e}$, which is proportional to $a$, then we need to 
change $M$. This is perhaps more easily appreciated in Appendix 
B of Hernquist (1990), where all the formulas for the dispersion 
express it as a function of $M/a$.
Figure 9 shows, for illustrative purposes only, the result of fitting 
the observed losvd with a Hernquist model. Clearly the NMAGIC modeling
by De Lorenzi et al. is to be preferred, but it is useful to show 
that the Hernquist model is not far off; the total mass of this
Hernquist nodel is 1.5$\times 10^{11} M_\odot$, which implies 
$(M/L)_B = 9$. This differs from the NMAGIC solution by less than 
20\% in the mass.

Let us go back to Figure 10. We have plotted two escape velocity 
curves, corresponding to $(M/L)_B = 9$ and 5. For a $(M/L)_B$ 
ratio of 9 we do not find any unbound PNs, but if we decrease 
$(M/L)_B$ to 5, then a few PNs become unbound. It should be clear 
that this kind of diagram cannot {\it prove\/} the absence of dark 
matter. Since we are dealing with projected distances to the center,  
some PNs can in fact be located at larger distances, where the 
escape velocity is lower. It is also possible that some PNs 
have large tangential velocities, so that the total spatial 
velocity can exceed the escape velocity. 
 
In summary, for the larger $M/L$ ratio, no dark matter is 
necessary; but, for the smaller $M/L$ ratio, the presence of 
several unbound PNs requires the presence of a dark matter halo.
A curve based on the NMAGIC mass of NGC 4697 falls between the two 
curves we plotted, and therefore the evidence is inconclusive. 
We can only say that the dark matter halo around NGC 4697 
is less conspicuous than the one around NGC 5128.

What is needed now is to repeat the NMAGIC modeling globally, taking 
into account both the old FORS and new FOCAS PN velocities. This will
require considerable work by other researchers, and is outside the
scope of the present paper, which is limited to presenting the 
new information we have obtained, and exploring its possible
significance.


\section{Rotation}

In Paper I we reported that
the rotation curve of NGC 4697 appeared to drop beyond about 100
arcsec from the center. We can reinvestigate the
rotation at large angular distances from the center of NGC 4697 
using our FOCAS PN velocities, which extend further out and have 
smaller errors than FORS velocities. Figure 11 shows histograms
for PNs progressively more distant from the center. We show separate
histograms for PNs with negative and positive $x$ coordinates.
In the left panel there is a peak of positive relative radial 
velocities at negative $x$, but anyway the signal of rotation 
(most of the galaxy moving toward us at negative $x$, and away from 
us at positive $x$), is clear. The same sense of rotation is also 
present in the central panel. The right panel shows that, at $x$
coordinates larger than 300 arcsec in absolute value, the rotation 
is no longer evident. From Paper I and De Lorenzi et al. (2008a) we
know already that there is no rotation along the minor axis (that is 
to say, around the major axis).

From long-slit absorption-line studies (Binney et al. 1990, De Lorenzi 
et al. 2008a) we know how the stars are rotating. Figure 12 shows
that the PNs rotate in the same sense. We have selected PNs within
50 arcsec of the major axis, and divided them in eight groups along
the major axis, with limits at $-$300, $-$200, $-$100, 0, 100, 200 
and 300 arcsec. The two extreme groups turned out to have too few PNs, 
and therefore we removed the $y$ limit in those two cases. In order of
increasing average $x$, the numbers of PNs in each group are 14, 16, 
19, 20, 11, 20, 13 and 8. We averaged the PN radial velocities and 
plotted the averages as a function of the average $x$ of each group.
The drop in rotation velocity at large radii is clear. It seems that 
NGC 4697 is hardly rotating at all at $5 R_{\rm e}$. The rotational 
signal must be somewhat diluted, because we were forced to include 
PNs with rather large $y$ coordinates; but still the behavior of
rotation is obvious. Figures 11 and 12 confirm what we suspected 
in Paper I. In relation to this lack of outer rotation, it may be 
interesting to mention a recent study of the S0 galaxy NGC 1023 
(Noordermeer et al. 2008), which has revealed a similar drop in 
the outer rotation velocity.

Sufficiently massive dark matter halos are usually expected to 
generate significant rotation in the outskirts, as discussed by 
De Lorenzi et al. (2008a). Indeed, for example NGC 5128 shows clear 
evidence of outer rotation (Woodley et al. 2007). That this is not
the case in NGC 4697 should not be interpreted as necessarily
indicating the lack of a dark matter halo. Instead, it may be a 
problem of how the specific angular momentum is distributed in the 
system, independently of its potential well and, by this, its total 
gravitational mass. In the same way as with the interpretation of 
the losvd, the interpretation of rotation will require more careful
global modeling with NMAGIC or similar codes. 

For the moment, the conclusion we want to extract is that there 
appears to exist a variety of cases concerning how much 
angular momentum resides in the outskirts of elliptical 
galaxies. PNs are a good tool that will allow to collect angular 
momentum information for several other galaxies. We hope that the 
information derived from PN radial velocities will make 
it possible to explore if this variety in angular momentum 
distribution can be traced to different galaxy formation 
mechanisms involving mergers. 

\section{Summary of conclusions}

We have demonstrated how Subaru and FOCAS can be used to obtain accurate 
slitless radial velocities of extragalactic PNs. Working in two fields
in the outskirts of NGC 4697, we discovered and provided radial velocities 
for 218 PNs, of which 162 had been previously observed with VLT+FORS. 
Where possible, we compared the FOCAS radial velocities with those 
previously obtained with FORS, finding good agreement. We now have PN 
kinematic information extending out to 5 effective radii from the center 
of NGC 4697 (its effective radius is 66 arcsec). In particular, 
we have significantly improved the kinematic information at angular 
distances larger than 230 arcsec from the center of NGC 4697.

We have found the following: 
(1) the losvd at radii between 300 and 400 arcsec from the center 
of NGC 4697 is below 100 km s$^{-1}$, i.e. a bit lower than obtained 
from the much smaller FORS outer PN sample by De Lorenzi et al. 
(2008a). This makes it difficult to narrow down the range of acceptable 
dark matter models. At least we can say that the dark matter halo
around NGC 4697 is not as conspicuous as the one around NGC 5128. 
(2) the rotational signal, which is obvious close to the center, 
is no longer visible at about 5 effective radii from the center.
By comparison to NGC 5128, it seems that much less angular momentum 
resides in the outskirts of NGC 4697. We need similar kinematic 
information about more elliptical galaxies.

A more quantitative interpretation of these results 
concerning NGC 4697 will require careful global modeling 
of the full, enlarged, FOCAS+FORS PN data set, plus existing 
absorption-line kinematic data, using NMAGIC or similar codes, 
following procedures described by De Lorenzi et al. (2008a).

\acknowledgements
This work was supported by the National Science Foundation under 
Grant No. 0307489. It is a pleasure to acknowledge the help provided 
by the Subaru staff, in particular the support astronomers Youichi 
Ohyama, Takashi Hattori and Kentaro Aoki. We would like to thank 
Fabrizio Bernardi for his help using David Tholen's astrometric 
programs for the calculation of J2000 equatorial coordinates.
We also acknowledge comments by R. Saglia, O. Gerhard and 
J. Barnes, as well as some very useful remarks by an anonymous 
referee.

\clearpage

\begin{deluxetable}{lrlrc}
\tablecaption{Observations and calibrations \label{tbl-1}}
\tablewidth{0pt}
\tablehead{
\colhead{FOCAS Field} & \colhead{Configuration} &
\colhead{FOCAS number} &
\colhead{exp (s)} & \colhead{Air mass\tablenotemark{a}}}
\startdata
 NGC 7293 + mask & on-band    &  52385 &   200 & 1.88 \\
 NGC 7293 + mask & on + grism &  52387 &   300 & 1.83 \\
 Th-Ar + mask    & on + grism &  52391 &    10 & 1.77 \\
 NGC 4697 W      & off-band   &  52445 &   120 & 1.17 \\
 NGC 4697 W      & on-band    &  52447 &  1200 & 1.15 \\  
 NGC 4697 W      & on + grism &  52449 &  1800 & 1.12 \\  
 Th-Ar + mask    & on + grism &  52453 &    10 & 1.11 \\ 
 Th-Ar + mask    & on-band    &  52455 &     4 & 1.11 \\ 
 NGC 4697 W      & off-band   &  52457 &   120 & 1.11 \\  
 NGC 4697 W      & on-band    &  52459 &  1200 & 1.11 \\  
 NGC 4697 W      & on + grism &  52461 &  1800 & 1.12 \\ 
 Th-Ar + mask    & on + grism &  52465 &    10 & 1.14 \\ 
 Th-Ar + mask    & on-band    &  52467 &     4 & 1.14 \\ 
 NGC 4697 W      & off-band   &  52469 &   120 & 1.15 \\  
 NGC 4697 W      & on-band    &  52471 &  1200 & 1.17 \\  
 NGC 4697 W      & on + grism &  52473 &  1800 & 1.23 \\  
 NGC 4697 W      & off-band   &  52481 &   120 & 1.31 \\  
 NGC 4697 W      & on-band    &  52483 &  1200 & 1.36 \\  
 NGC 4697 W      & on + grism &  52485 &  1800 & 1.70 \\  
 NGC 4697 E      & off-band   &  60793 &   120 & 1.24 \\  
 NGC 4697 E      & on-band    &  60795 &  1200 & 1.21 \\  
 NGC 4697 E      & on + grism &  60797 &  1800 & 1.16 \\  
 NGC 4697 E      & off-band   &  60803 &   120 & 1.13 \\  
 NGC 4697 E      & on-band    &  60805 &  1200 & 1.12 \\  
 NGC 4697 E      & on + grism &  60807 &  1800 & 1.11 \\  
 NGC 4697 E      & off-band   &  60813 &   120 & 1.12 \\  
 NGC 4697 E      & on-band    &  60815 &  1200 & 1.13 \\  
 NGC 4697 E      & on + grism &  60817 &  1800 & 1.16 \\  
 NGC 4697 E      & off-band   &  60823 &   120 & 1.22 \\  
 NGC 4697 E      & on-band    &  60825 &  1200 & 1.26 \\  
 NGC 4697 E      & on + grism &  60827 &  1800 & 1.35 \\  
 NGC 4697 E      & on + grism &  60833 &  2000 & 1.59 \\  
 G 138-31        & on-band    &  60843 &    60 & 1.02 \\
 G 138-31        & on-band    &  60845 &   120 & 1.02 \\
 G 138-31        & on-band    &  60847 &    60 & 1.02 \\
 G 138-31        & on-band    &  60849 &    60 & 1.02 \\
 G 138-31        & on-band    &  60851 &   120 & 1.02 \\
 G 138-31        & on-band    &  60879 &    60 & 1.34 \\
 G 138-31        & on-band    &  60881 &   120 & 1.36 \\
 G 138-31        & on-band    &  60883 &    60 & 1.38 \\
 G 138-31        & on-band    &  60885 &   120 & 1.40 \\
 NGC 4697 W      & off-band   &  60967 &   120 & 1.24 \\  
 NGC 4697 W      & on-band    &  60969 &  1200 & 1.21 \\  
 NGC 4697 W      & on + grism &  60971 &  1800 & 1.16 \\  
 NGC 4697 W      & off-band   &  60977 &   120 & 1.12 \\  
 NGC 4697 W      & on-band    &  60979 &  1200 & 1.12 \\  
 NGC 4697 W      & on + grism &  60981 &  1800 & 1.11 \\  
 NGC 4697 E      & off-band   &  60987 &   120 & 1.12 \\  
 NGC 4697 E      & on-band    &  60989 &  1200 & 1.13 \\  
 NGC 4697 E      & on + grism &  60991 &  1800 & 1.16 \\  
 Th-Ar + mask    & on + grism &  60993 &    10 & 1.20 \\ 
 Th-Ar + mask    & on-band    &  60995 &     4 & 1.20 \\ 
 NGC 4697 E      & off-band   &  60997 &   120 & 1.21 \\  
 NGC 4697 E      & on-band    &  60999 &  1200 & 1.24 \\  
 NGC 4697 E      & on + grism &  61001 &  1800 & 1.35 \\  
 NGC 4697 W      & off-band   &  61011 &   120 & 1.55 \\  
 NGC 4697 W      & on-band    &  61013 &  1200 & 1.65 \\  
 NGC 4697 W      & on + grism &  61015 &  2000 & 1.96 \\  
 G 138-31        & on-band    &  61017 &   120 & 1.02 \\
 G 138-31        & on-band    &  61019 &   120 & 1.02 \\
 G 138-31        & on-band    &  61021 &   120 & 1.02 \\
 G 138-31        & on-band    &  61023 &   120 & 1.03 \\
 NGC 4697 E      & off-band   &  61081 &   140 & 1.24 \\  
 NGC 4697 E      & on-band    &  61083 &  1400 & 1.20 \\  
 NGC 4697 E      & on + grism &  61085 &  2200 & 1.15 \\  
 NGC 4697 E      & off-band   &  61091 &   140 & 1.12 \\  
 NGC 4697 E      & on-band    &  61093 &  1400 & 1.11 \\  
 NGC 4697 E      & on + grism &  61095 &  2200 & 1.12 \\  
 NGC 4697 W      & off-band   &  61101 &   140 & 1.13 \\  
 NGC 4697 W      & on-band    &  61103 &  1400 & 1.15 \\  
 NGC 4697 W      & on + grism &  61105 &  2200 & 1.22 \\  
 NGC 4697 W      & off-band   &  61111 &   140 & 1.30 \\  
 NGC 4697 W      & on-band    &  61113 &  1400 & 1.36 \\  
 NGC 4697 W      & on + grism &  61115 &  2200 & 1.54 \\  
 NGC 4697 W      & on + grism &  61119 &  2200 & 1.93 \\  
\enddata
\tablenotetext{a}{the air masses correspond to the middle of each exposure}
\end{deluxetable}

\clearpage

\begin{deluxetable}{rrrrrrrrrrrrr}
\tablecaption{Planetary nebulae in NGC 4697 detected with FOCAS \tablenotemark{a} \label{tbl-2}}
\tablewidth{0pt}
\rotate
\tabletypesize{\tiny}
\tablehead{
\colhead{Id, F} & \colhead{$x$, G} & \colhead{$y$, G} & 
\colhead{\ } & \colhead{$\alpha$} & \colhead{(2000)} & 
\colhead{\ } & \colhead{$\delta$} & \colhead{(2000)} & \colhead{$m(5007)$} &
\colhead{Helioc. RV} & \colhead{Id, E} & \colhead{Id, W}
} 
\startdata

   2001 &    376 &     -2 &   12 &   48 &   13.13 &   -5 &   50 &   43.3 &    27.02 &  1256 &      -1 &    -1 \\
   2002 &    378 &   -128 &   12 &   48 &   16.59 &   -5 &   52 &   38.1 &    27.01 &  1240 &      -1 &    -1 \\
   2003 &    386 &   -129 &   12 &   48 &   16.12 &   -5 &   52 &   42.5 &    26.98 &  1282 &      -1 &    -1 \\
   2004 &    287 &   -191 &   12 &   48 &   23.89 &   -5 &   52 &   56.8 &    26.67 &  1314 &      -1 &    -1 \\
   2005 &    312 &    -16 &   12 &   48 &   17.45 &   -5 &   50 &   28.9 &    27.84 &  1134 &      -1 &    -1 \\
   2006 &    338 &   -142 &   12 &   48 &   19.37 &   -5 &   52 &   34.1 &    28.33 &  1362 &      -1 &    -1 \\
   2007 &    286 &    -13 &   12 &   48 &   18.91 &   -5 &   50 &   15.3 &    25.90 &  1367 &      -1 &  1746 \\
   2009 &    281 &    -69 &   12 &   48 &   20.79 &   -5 &   51 &    3.5 &    26.49 &  1178 &      -1 &  1742 \\
   2010 &    278 &    -94 &   12 &   48 &   21.71 &   -5 &   51 &   25.4 &    27.51 &  1400 &      -1 &    -1 \\
   2011 &    281 &   -136 &   12 &   48 &   22.70 &   -5 &   52 &    4.1 &    27.60 &  1251 &      -1 &    -1 \\
   2012 &    260 &   -155 &   12 &   48 &   24.48 &   -5 &   52 &   12.9 &    27.90 &  1346 &      -1 &    -1 \\
   2013 &    250 &   -130 &   12 &   48 &   24.40 &   -5 &   51 &   45.7 &    28.06 &  1326 &      -1 &    -1 \\
   2014 &    236 &    -80 &   12 &   48 &   23.82 &   -5 &   50 &   54.6 &    26.47 &  1218 &      -1 &  1734 \\
   2015 &    260 &    -61 &   12 &   48 &   21.84 &   -5 &   50 &   48.0 &    26.95 &  1370 &      -1 &  1743 \\
   2016 &    269 &    -34 &   12 &   48 &   20.52 &   -5 &   50 &   27.2 &    26.93 &  1312 &      -1 &  1745 \\
   2017 &    271 &    -44 &   12 &   48 &   20.71 &   -5 &   50 &   36.5 &    -1.   &  1450 &      -1 &  1744 \\
   2018 &    239 &     -8 &   12 &   48 &   21.59 &   -5 &   49 &   50.9 &    27.43 &  1208 &      -1 &    -1 \\
   2019 &    236 &     -5 &   12 &   48 &   21.71 &   -5 &   49 &   46.5 &    27.97 &  1309 &      -1 &    -1 \\
   2020 &    224 &    -38 &   12 &   48 &   23.38 &   -5 &   50 &   11.6 &    27.09 &  1472 &      -1 &  1736 \\
   2021 &    216 &    -28 &   12 &   48 &   23.62 &   -5 &   49 &   59.0 &    25.97 &  1332 &      -1 &  1737 \\
   2022 &    211 &    -45 &   12 &   48 &   24.37 &   -5 &   50 &   12.8 &    27.19 &  1280 &      -1 &  1735 \\
   2024 &    235 &    -92 &   12 &   48 &   24.26 &   -5 &   51 &    4.5 &    27.16 &  1368 &      -1 &  1733 \\
   2025 &    220 &   -137 &   12 &   48 &   26.43 &   -5 &   51 &   40.1 &    26.03 &  1265 &      -1 &  1732 \\
   2026 &    196 &   -150 &   12 &   48 &   28.23 &   -5 &   51 &   41.1 &    27.51 &  1312 &      -1 &    -1 \\
   2027 &    191 &    -80 &   12 &   48 &   26.55 &   -5 &   50 &   35.9 &    27.23 &  1135 &      -1 &  1722 \\
   2028 &    202 &    -45 &   12 &   48 &   24.90 &   -5 &   50 &    9.0 &    26.05 &  1315 &      -1 &  1723 \\
   2029 &    187 &    -16 &   12 &   48 &   25.01 &   -5 &   49 &   36.3 &    27.57 &  1399 &      -1 &    -1 \\
   2030 &    184 &    -51 &   12 &   48 &   26.20 &   -5 &   50 &    6.1 &    27.10 &  1322 &      -1 &    -1 \\
   2031 &    165 &   -120 &   12 &   48 &   29.26 &   -5 &   51 &    0.8 &    26.56 &  1286 &      -1 &  1700 \\
   2032 &    130 &    -27 &   12 &   48 &   28.79 &   -5 &   49 &   21.4 &    26.15 &  1430 &      -1 &  1668 \\
   2033 &    153 &    -21 &   12 &   48 &   27.18 &   -5 &   49 &   25.8 &    25.96 &  1234 &      -1 &  1707 \\
   2034 &    141 &    -65 &   12 &   48 &   29.19 &   -5 &   50 &    1.0 &    26.16 &  1097 &      -1 &  1703 \\
   2035 &    126 &    -78 &   12 &   48 &   30.46 &   -5 &   50 &    6.5 &    25.64 &  1243 &      -1 &  1661 \\
   2036 &    108 &    -38 &   12 &   48 &   30.41 &   -5 &   49 &   22.8 &    26.51 &  1193 &      -1 &  1663 \\
   2037 &    198 &    -13 &   12 &   48 &   24.25 &   -5 &   49 &   38.3 &    27.44 &  1517 &      -1 &    -1 \\
   2038 &     74 &    -25 &   12 &   48 &   32.14 &   -5 &   48 &   55.9 &    25.72 &  1348 &      -1 &  1627 \\
   2039 &    108 &    -10 &   12 &   48 &   29.62 &   -5 &   48 &   57.6 &    26.29 &  1365 &      -1 &  1666 \\
   2040 &     92 &    -27 &   12 &   48 &   31.10 &   -5 &   49 &    5.3 &    26.85 &  1400 &      -1 &  1664 \\
   2041 &    126 &    -21 &   12 &   48 &   28.84 &   -5 &   49 &   15.1 &    26.63 &  1464 &      -1 &  1667 \\
   2042 &     85 &    -51 &   12 &   48 &   32.21 &   -5 &   49 &   24.5 &    26.20 &   953 &      -1 &  1623 \\
   2043 &    120 &    -54 &   12 &   48 &   30.15 &   -5 &   49 &   41.7 &    27.67 &  1449 &      -1 &    -1 \\
   2044 &    104 &    -54 &   12 &   48 &   31.09 &   -5 &   49 &   35.0 &    27.08 &  1425 &      -1 &  1662 \\
   2045 &    120 &   -103 &   12 &   48 &   31.52 &   -5 &   50 &   26.9 &    27.30 &  1101 &      -1 &  1660 \\
   2046 &    147 &    -84 &   12 &   48 &   29.34 &   -5 &   50 &   21.2 &    26.78 &  1312 &      -1 &  1701 \\
   2047 &    133 &    -78 &   12 &   48 &   30.01 &   -5 &   50 &    9.1 &    26.85 &  1222 &      -1 &  1702 \\
   2048 &    156 &    -72 &   12 &   48 &   28.48 &   -5 &   50 &   13.5 &    27.67 &  1245 &      -1 &    -1 \\
   2050 &     69 &    -13 &   12 &   48 &   32.10 &   -5 &   48 &   43.1 &    -1.   &  1407 &      -1 &  1628 \\
   2051 &     81 &    -14 &   12 &   48 &   31.38 &   -5 &   48 &   49.7 &    26.77 &  1388 &      -1 &  1630 \\
   2101 &    273 &    112 &   12 &   48 &   16.16 &   -5 &   48 &   15.8 &    26.93 &  1261 &      -1 &    -1 \\
   2102 &    270 &    102 &   12 &   48 &   16.62 &   -5 &   48 &   23.9 &    27.53 &  1207 &      -1 &    -1 \\
   2103 &    279 &    100 &   12 &   48 &   16.16 &   -5 &   48 &   29.7 &    27.12 &  1540 &      -1 &    -1 \\
   2104 &    329 &     89 &   12 &   48 &   13.42 &   -5 &   49 &    0.2 &    27.61 &  1375 &      -1 &    -1 \\
   2105 &    313 &     60 &   12 &   48 &   15.18 &   -5 &   49 &   20.5 &    26.30 &  1270 &      -1 &    -1 \\
   2106 &    356 &    162 &   12 &   48 &    9.72 &   -5 &   48 &    6.1 &    27.15 &  1276 &      -1 &    -1 \\
   2107 &    281 &    164 &   12 &   48 &   14.18 &   -5 &   47 &   32.2 &    28.16 &  1335 &      -1 &    -1 \\
   2108 &    278 &     30 &   12 &   48 &   18.18 &   -5 &   49 &   33.1 &    28.45 &  1329 &      -1 &    -1 \\
   2109 &    241 &     66 &   12 &   48 &   19.41 &   -5 &   48 &   44.2 &    26.92 &  1380 &      -1 &  1738 \\
   2110 &    225 &     69 &   12 &   48 &   20.30 &   -5 &   48 &   34.7 &    27.67 &  1339 &      -1 &  1739 \\
   2111 &    211 &     79 &   12 &   48 &   20.87 &   -5 &   48 &   19.9 &    -1.   &  1185 &      -1 &  1740 \\
   2112 &    199 &     67 &   12 &   48 &   21.94 &   -5 &   48 &   25.4 &    26.23 &  1359 &      -1 &  1728 \\
   2113 &    207 &     36 &   12 &   48 &   22.31 &   -5 &   48 &   57.3 &    26.08 &  1290 &      -1 &  1727 \\
   2114 &    207 &     26 &   12 &   48 &   22.63 &   -5 &   49 &    6.2 &    27.40 &  1205 &      -1 &  1725 \\
   2115 &    194 &     31 &   12 &   48 &   23.24 &   -5 &   48 &   55.9 &    27.65 &  1403 &      -1 &  1726 \\
   2116 &    203 &     15 &   12 &   48 &   23.14 &   -5 &   49 &   14.5 &    27.44 &  1125 &      -1 &  1724 \\
   2117 &    186 &     45 &   12 &   48 &   23.35 &   -5 &   48 &   40.5 &    27.68 &  1316 &      -1 &    -1 \\
   2118 &    157 &     37 &   12 &   48 &   25.32 &   -5 &   48 &   35.7 &    27.04 &  1349 &      -1 &  1714 \\
   2119 &    165 &     41 &   12 &   48 &   24.69 &   -5 &   48 &   34.9 &    26.48 &  1204 &      -1 &  1715 \\
   2120 &    175 &    108 &   12 &   48 &   22.20 &   -5 &   47 &   38.6 &    27.30 &  1511 &      -1 &  1729 \\
   2121 &    168 &     85 &   12 &   48 &   23.29 &   -5 &   47 &   56.0 &    27.27 &  1339 &      -1 &  1717 \\
   2122 &    159 &     66 &   12 &   48 &   24.39 &   -5 &   48 &    9.3 &    27.88 &  1437 &      -1 &  1716 \\
   2123 &    125 &     72 &   12 &   48 &   26.25 &   -5 &   47 &   49.7 &    26.81 &  1401 &      -1 &  1686 \\
   2124 &    173 &     70 &   12 &   48 &   23.38 &   -5 &   48 &   11.8 &    27.99 &  1323 &      -1 &    -1 \\
   2125 &    114 &    116 &   12 &   48 &   25.67 &   -5 &   47 &    5.6 &    26.39 &  1357 &      -1 &  1692 \\
   2126 &    110 &     86 &   12 &   48 &   26.80 &   -5 &   47 &   30.9 &    27.68 &  1076 &      -1 &  1688 \\
   2127 &    121 &     85 &   12 &   48 &   26.16 &   -5 &   47 &   36.5 &    28.00 &  1554 &      -1 &  1689 \\
   2128 &     97 &     96 &   12 &   48 &   27.27 &   -5 &   47 &   16.3 &    26.80 &  1268 &      -1 &  1691 \\
   2129 &     99 &     82 &   12 &   48 &   27.58 &   -5 &   47 &   29.7 &    26.48 &  1286 &      -1 &  1687 \\
   2130 &    100 &     75 &   12 &   48 &   27.73 &   -5 &   47 &   36.4 &    27.05 &  1290 &      -1 &  1685 \\
   2131 &     94 &     70 &   12 &   48 &   28.23 &   -5 &   47 &   38.2 &    26.63 &  1415 &      -1 &  1684 \\
   2132 &    149 &    142 &   12 &   48 &   22.85 &   -5 &   46 &   56.8 &    27.51 &  1301 &      -1 &  1718 \\
   2133 &    152 &    138 &   12 &   48 &   22.75 &   -5 &   47 &    1.2 &    27.57 &  1471 &      -1 &    -1 \\
   2134 &    194 &    133 &   12 &   48 &   20.33 &   -5 &   47 &   24.0 &    27.62 &  1265 &      -1 &  1730 \\
   2135 &    138 &     42 &   12 &   48 &   26.32 &   -5 &   48 &   22.3 &    26.31 &  1064 &      -1 &  1713 \\
   2136 &    136 &     33 &   12 &   48 &   26.70 &   -5 &   48 &   29.5 &    26.28 &  1175 &      -1 &  1711 \\
   2137 &    133 &     34 &   12 &   48 &   26.86 &   -5 &   48 &   27.6 &    26.44 &  1063 &      -1 &  1712 \\
   2138 &    122 &     31 &   12 &   48 &   27.63 &   -5 &   48 &   25.5 &    27.08 &  1323 &      -1 &  1677 \\
   2139 &    117 &     59 &   12 &   48 &   27.13 &   -5 &   47 &   58.8 &    25.63 &  1095 &      -1 &  1683 \\
   2140 &    109 &     59 &   12 &   48 &   27.60 &   -5 &   47 &   55.6 &    26.59 &  1240 &      -1 &  1682 \\
   2141 &    135 &     14 &   12 &   48 &   27.31 &   -5 &   48 &   46.8 &    27.33 &  1321 &      -1 &    -1 \\
   2142 &    135 &      2 &   12 &   48 &   27.67 &   -5 &   48 &   58.0 &    27.00 &  1462 &      -1 &  1710 \\
   2143 &     85 &     61 &   12 &   48 &   28.98 &   -5 &   47 &   42.9 &    27.03 &  1410 &      -1 &    -1 \\
   2144 &     95 &     53 &   12 &   48 &   28.62 &   -5 &   47 &   54.5 &    26.51 &  1087 &      -1 &  1681 \\
   2145 &     99 &     42 &   12 &   48 &   28.67 &   -5 &   48 &    6.1 &    26.69 &  1315 &      -1 &  1679 \\
   2146 &     95 &     42 &   12 &   48 &   28.95 &   -5 &   48 &    5.0 &    26.71 &  1351 &      -1 &  1678 \\
   2147 &     98 &     50 &   12 &   48 &   28.55 &   -5 &   47 &   58.4 &    27.55 &  1287 &      -1 &  1680 \\
   2148 &     87 &     35 &   12 &   48 &   29.64 &   -5 &   48 &    7.7 &    27.22 &  1222 &      -1 &  1644 \\
   2149 &    112 &      2 &   12 &   48 &   29.02 &   -5 &   48 &   48.2 &    26.08 &  1303 &      -1 &  1671 \\
   2150 &    109 &      8 &   12 &   48 &   29.04 &   -5 &   48 &   41.4 &    26.70 &  1180 &      -1 &  1673 \\
   2151 &    104 &     16 &   12 &   48 &   29.15 &   -5 &   48 &   31.3 &    26.53 &  1377 &      -1 &  1676 \\
   2152 &     78 &     30 &   12 &   48 &   30.33 &   -5 &   48 &    7.8 &    26.61 &  1160 &      -1 &  1643 \\
   2153 &     85 &     27 &   12 &   48 &   29.97 &   -5 &   48 &   13.7 &    27.20 &  1170 &      -1 &  1642 \\
   2154 &     86 &     19 &   12 &   48 &   30.11 &   -5 &   48 &   21.2 &    27.28 &  1060 &      -1 &  1641 \\
   2155 &     74 &      3 &   12 &   48 &   31.32 &   -5 &   48 &   30.8 &    25.87 &  1414 &      -1 &  1636 \\
   2201 &   -265 &   -109 &   12 &   48 &   55.10 &   -5 &   47 &   48.8 &    27.77 &  1107 &      27 &    -1 \\
   2202 &   -393 &   -128 &   12 &   49 &    3.37 &   -5 &   47 &   12.2 &    27.07 &  1297 &      -1 &    -1 \\
   2203 &   -340 &   -157 &   12 &   49 &    0.98 &   -5 &   48 &    0.7 &    26.23 &  1257 &      -1 &    -1 \\
   2204 &   -349 &    -51 &   12 &   48 &   58.55 &   -5 &   46 &   21.6 &    25.70 &  1289 &      -1 &    -1 \\
   2205 &   -342 &    -46 &   12 &   48 &   57.98 &   -5 &   46 &   19.6 &    27.24 &  1349 &      -1 &    -1 \\
   2206 &   -331 &    -57 &   12 &   48 &   57.60 &   -5 &   46 &   34.6 &    26.64 &  1244 &      -1 &    -1 \\
   2207 &   -318 &    -54 &   12 &   48 &   56.76 &   -5 &   46 &   37.0 &    26.74 &  1173 &      -1 &    -1 \\
   2208 &   -258 &    -38 &   12 &   48 &   52.63 &   -5 &   46 &   47.8 &    26.55 &  1220 &      65 &    -1 \\
   2209 &   -251 &    -53 &   12 &   48 &   52.66 &   -5 &   47 &    4.0 &    26.25 &  1153 &      66 &    -1 \\
   2210 &   -244 &    -52 &   12 &   48 &   52.16 &   -5 &   47 &    6.2 &    26.88 &  1189 &      67 &    -1 \\
   2211 &   -242 &    -47 &   12 &   48 &   51.96 &   -5 &   47 &    2.8 &    27.78 &  1094 &      68 &    -1 \\
   2212 &   -228 &    -26 &   12 &   48 &   50.48 &   -5 &   46 &   49.4 &    27.77 &  1314 &      -1 &    -1 \\
   2213 &   -338 &    -16 &   12 &   48 &   56.88 &   -5 &   45 &   54.5 &    26.12 &  1261 &      -1 &    -1 \\
   2214 &   -303 &    -28 &   12 &   48 &   55.06 &   -5 &   46 &   19.5 &    26.72 &  1310 &     114 &    -1 \\
   2215 &   -221 &   -113 &   12 &   48 &   52.56 &   -5 &   48 &   11.3 &    28.09 &  1245 &      -1 &    -1 \\
   2216 &   -261 &   -131 &   12 &   48 &   55.45 &   -5 &   48 &   10.9 &    27.55 &  1167 &      11 &    -1 \\
   2217 &   -255 &   -167 &   12 &   48 &   56.13 &   -5 &   48 &   45.6 &    26.73 &  1085 &       2 &    -1 \\
   2218 &   -294 &   -179 &   12 &   48 &   58.82 &   -5 &   48 &   40.5 &    26.65 &  1363 &       1 &    -1 \\
   2219 &   -215 &   -157 &   12 &   48 &   53.42 &   -5 &   48 &   53.4 &    27.60 &  1289 &       3 &    -1 \\
   2220 &   -217 &    -33 &   12 &   48 &   50.00 &   -5 &   47 &    0.7 &    26.78 &  1189 &      69 &    -1 \\
   2221 &   -209 &    -34 &   12 &   48 &   49.53 &   -5 &   47 &    5.0 &    -1.   &  1224 &      70 &    -1 \\
   2222 &   -201 &    -13 &   12 &   48 &   48.49 &   -5 &   46 &   49.1 &    27.06 &  1356 &     118 &    -1 \\
   2223 &   -178 &    -30 &   12 &   48 &   47.59 &   -5 &   47 &   14.1 &    27.03 &  1388 &     121 &    -1 \\
   2224 &   -172 &    -18 &   12 &   48 &   46.88 &   -5 &   47 &    6.3 &    27.12 &  1297 &     120 &    -1 \\
   2225 &   -143 &     -9 &   12 &   48 &   44.83 &   -5 &   47 &   10.4 &    27.43 &  1051 &     128 &    -1 \\
   2226 &   -150 &    -38 &   12 &   48 &   46.12 &   -5 &   47 &   33.8 &    27.45 &  1124 &      73 &    -1 \\
   2227 &   -124 &    -11 &   12 &   48 &   43.76 &   -5 &   47 &   20.3 &    27.17 &  1245 &     129 &    -1 \\
   2228 &   -107 &    -14 &   12 &   48 &   42.80 &   -5 &   47 &   29.4 &    25.93 &  1495 &     130 &    -1 \\
   2229 &    -95 &    -29 &   12 &   48 &   42.46 &   -5 &   47 &   48.5 &    26.35 &  1409 &     142 &    -1 \\
   2230 &    -83 &    -23 &   12 &   48 &   41.59 &   -5 &   47 &   47.8 &    26.51 &  1074 &     140 &    -1 \\
   2231 &    -86 &    -23 &   12 &   48 &   41.77 &   -5 &   47 &   46.9 &    -1.   &  1190 &     139 &    -1 \\
   2232 &    -85 &    -25 &   12 &   48 &   41.77 &   -5 &   47 &   49.3 &    27.36 &  1486 &     141 &    -1 \\
   2233 &    -89 &    -11 &   12 &   48 &   41.65 &   -5 &   47 &   35.0 &    27.50 &   980 &     137 &    -1 \\
   2234 &    -79 &    -14 &   12 &   48 &   41.09 &   -5 &   47 &   41.8 &    26.73 &  1474 &     138 &    -1 \\
   2235 &    -76 &    -47 &   12 &   48 &   41.87 &   -5 &   48 &   12.8 &    26.56 &  1015 &      82 &    -1 \\
   2236 &    -79 &    -43 &   12 &   48 &   41.94 &   -5 &   48 &    7.7 &    27.31 &  1115 &      81 &    -1 \\
   2237 &    -88 &    -58 &   12 &   48 &   42.90 &   -5 &   48 &   17.3 &    26.89 &  1164 &      83 &    -1 \\
   2238 &    -99 &    -67 &   12 &   48 &   43.83 &   -5 &   48 &   21.1 &    26.54 &  1249 &      84 &    -1 \\
   2239 &   -118 &    -69 &   12 &   48 &   45.03 &   -5 &   48 &   15.4 &    26.83 &  1304 &      78 &    -1 \\
   2240 &   -144 &    -80 &   12 &   48 &   46.93 &   -5 &   48 &   13.8 &    25.82 &  1165 &      36 &    -1 \\
   2241 &   -132 &   -107 &   12 &   48 &   46.92 &   -5 &   48 &   43.6 &    26.21 &  1016 &      41 &    -1 \\
   2242 &   -126 &   -106 &   12 &   48 &   46.59 &   -5 &   48 &   45.5 &    26.96 &  1333 &      42 &    -1 \\
   2243 &   -110 &   -113 &   12 &   48 &   45.77 &   -5 &   48 &   57.9 &    26.50 &  1250 &      15 &    -1 \\
   2244 &   -137 &    -87 &   12 &   48 &   46.71 &   -5 &   48 &   23.3 &    27.48 &  1170 &      37 &    -1 \\
   2245 &   -113 &    -85 &   12 &   48 &   45.19 &   -5 &   48 &   32.1 &    27.26 &  1019 &      40 &    -1 \\
   2246 &   -120 &    -82 &   12 &   48 &   45.49 &   -5 &   48 &   26.4 &    27.63 &  1068 &      39 &    -1 \\
   2247 &   -139 &    -71 &   12 &   48 &   46.37 &   -5 &   48 &    8.4 &    27.98 &  1150 &      -1 &    -1 \\
   2248 &   -155 &    -95 &   12 &   48 &   47.99 &   -5 &   48 &   23.1 &    26.99 &  1246 &      35 &    -1 \\
   2249 &   -157 &    -89 &   12 &   48 &   47.96 &   -5 &   48 &   17.0 &    27.51 &  1283 &      34 &    -1 \\
   2250 &   -166 &   -106 &   12 &   48 &   49.02 &   -5 &   48 &   28.6 &    26.51 &  1159 &      31 &    -1 \\
   2251 &   -169 &   -109 &   12 &   48 &   49.25 &   -5 &   48 &   29.5 &    27.52 &  1076 &      30 &    -1 \\
   2252 &   -162 &   -108 &   12 &   48 &   48.80 &   -5 &   48 &   32.2 &    27.52 &  1307 &      32 &    -1 \\
   2253 &   -172 &   -133 &   12 &   48 &   50.11 &   -5 &   48 &   50.2 &    27.06 &  1343 &      12 &    -1 \\
   2254 &   -133 &   -123 &   12 &   48 &   47.48 &   -5 &   48 &   57.3 &    -1.   &  1241 &      13 &    -1 \\
   2255 &   -198 &    -81 &   12 &   48 &   50.21 &   -5 &   47 &   51.8 &    26.60 &  1351 &      28 &    -1 \\
   2256 &   -175 &   -101 &   12 &   48 &   49.38 &   -5 &   48 &   20.0 &    27.40 &  1214 &      29 &    -1 \\
   2257 &   -179 &    -64 &   12 &   48 &   48.58 &   -5 &   47 &   44.9 &    27.98 &  1139 &      71 &    -1 \\
   2258 &   -111 &    -43 &   12 &   48 &   43.89 &   -5 &   47 &   54.8 &    27.47 &  1176 &      76 &    -1 \\
   2259 &   -103 &    -38 &   12 &   48 &   43.24 &   -5 &   47 &   53.0 &    27.34 &  1096 &      79 &    -1 \\
   2260 &    -66 &     -6 &   12 &   48 &   40.07 &   -5 &   47 &   39.9 &    -1.   &  1140 &     136 &  1024 \\
   2261 &    -59 &    -15 &   12 &   48 &   39.92 &   -5 &   47 &   50.5 &    26.61 &  1207 &     162 &  1043 \\
   2262 &    -61 &    -14 &   12 &   48 &   40.00 &   -5 &   47 &   49.3 &    -1.   &  1123 &     161 &  1042 \\
   2301 &    -88 &     71 &   12 &   48 &   39.25 &   -5 &   46 &   20.6 &    26.61 &  1319 &     300 &    -1 \\
   2302 &    -89 &     50 &   12 &   48 &   39.92 &   -5 &   46 &   39.0 &    26.73 &  1218 &     301 &    -1 \\
   2303 &    -92 &     45 &   12 &   48 &   40.22 &   -5 &   46 &   43.3 &    27.45 &  1148 &     220 &    -1 \\
   2304 &   -105 &     31 &   12 &   48 &   41.39 &   -5 &   46 &   49.7 &    26.78 &  1402 &     219 &    -1 \\
   2305 &    -86 &     29 &   12 &   48 &   40.34 &   -5 &   47 &    0.1 &    26.95 &  1332 &     222 &    -1 \\
   2306 &    -94 &     20 &   12 &   48 &   41.09 &   -5 &   47 &    4.7 &    27.05 &  1341 &     221 &    -1 \\
   2307 &    -74 &     24 &   12 &   48 &   39.72 &   -5 &   47 &    9.5 &    26.55 &   973 &     226 &    -1 \\
   2308 &    -76 &     25 &   12 &   48 &   39.79 &   -5 &   47 &    7.6 &    27.24 &  1220 &     225 &    -1 \\
   2309 &    -70 &     32 &   12 &   48 &   39.22 &   -5 &   47 &    4.0 &    27.15 &  1254 &     228 &    -1 \\
   2310 &    -63 &     34 &   12 &   48 &   38.78 &   -5 &   47 &    5.1 &    26.55 &  1069 &     229 &  1050 \\
   2311 &   -117 &    129 &   12 &   48 &   39.34 &   -5 &   45 &   16.5 &    26.68 &  1157 &     362 &    -1 \\
   2312 &   -129 &    127 &   12 &   48 &   40.17 &   -5 &   45 &   13.1 &    26.27 &  1324 &     343 &    -1 \\
   2313 &   -137 &    119 &   12 &   48 &   40.85 &   -5 &   45 &   16.3 &    27.30 &  1389 &     342 &    -1 \\
   2314 &   -113 &     96 &   12 &   48 &   40.07 &   -5 &   45 &   47.4 &    27.15 &  1011 &     346 &    -1 \\
   2315 &   -112 &     64 &   12 &   48 &   40.87 &   -5 &   46 &   16.8 &    26.78 &  1124 &     298 &    -1 \\
   2316 &   -125 &     28 &   12 &   48 &   42.74 &   -5 &   46 &   44.5 &    26.38 &  1161 &     217 &    -1 \\
   2317 &   -130 &     26 &   12 &   48 &   43.04 &   -5 &   46 &   44.3 &    -1.   &  1320 &     216 &    -1 \\
   2318 &   -124 &      6 &   12 &   48 &   43.26 &   -5 &   47 &    4.3 &    26.88 &  1035 &     126 &    -1 \\
   2319 &   -156 &     19 &   12 &   48 &   44.83 &   -5 &   46 &   39.8 &    27.38 &  1375 &      -1 &    -1 \\
   2320 &   -172 &      7 &   12 &   48 &   46.13 &   -5 &   46 &   43.4 &    26.03 &  1313 &     119 &    -1 \\
   2321 &   -168 &     63 &   12 &   48 &   44.32 &   -5 &   45 &   54.5 &    26.54 &  1464 &     295 &    -1 \\
   2322 &   -149 &     80 &   12 &   48 &   42.68 &   -5 &   45 &   46.5 &    27.47 &  1211 &     296 &    -1 \\
   2323 &   -146 &    102 &   12 &   48 &   41.90 &   -5 &   45 &   28.3 &    28.22 &  1336 &     340 &    -1 \\
   2324 &   -156 &     99 &   12 &   48 &   42.55 &   -5 &   45 &   26.4 &    27.44 &  1279 &     339 &    -1 \\
   2325 &   -151 &     93 &   12 &   48 &   42.41 &   -5 &   45 &   34.0 &    27.87 &  1327 &     341 &    -1 \\
   2326 &   -166 &    125 &   12 &   48 &   42.47 &   -5 &   44 &   58.7 &    27.79 &  1349 &     337 &    -1 \\
   2327 &   -181 &     19 &   12 &   48 &   46.37 &   -5 &   46 &   28.7 &    26.68 &  1030 &      -1 &    -1 \\
   2328 &   -183 &     13 &   12 &   48 &   46.65 &   -5 &   46 &   33.3 &    27.71 &  1193 &      -1 &    -1 \\
   2329 &   -188 &     27 &   12 &   48 &   46.55 &   -5 &   46 &   18.9 &    26.42 &  1331 &      -1 &    -1 \\
   2330 &   -188 &     26 &   12 &   48 &   46.57 &   -5 &   46 &   19.5 &    26.30 &  1223 &      -1 &    -1 \\
   2331 &   -189 &     31 &   12 &   48 &   46.53 &   -5 &   46 &   14.7 &    27.69 &  1251 &      -1 &    -1 \\
   2332 &   -201 &     17 &   12 &   48 &   47.64 &   -5 &   46 &   21.6 &    26.90 &  1037 &     214 &    -1 \\
   2333 &   -206 &     19 &   12 &   48 &   47.86 &   -5 &   46 &   18.4 &    26.85 &  1275 &     213 &    -1 \\
   2334 &   -216 &     12 &   12 &   48 &   48.70 &   -5 &   46 &   20.2 &    26.84 &  1019 &     212 &    -1 \\
   2335 &   -223 &      6 &   12 &   48 &   49.27 &   -5 &   46 &   22.5 &    26.64 &  1316 &     116 &    -1 \\
   2336 &   -204 &     77 &   12 &   48 &   46.11 &   -5 &   45 &   25.9 &    27.66 &  1148 &     293 &    -1 \\
   2337 &   -252 &     16 &   12 &   48 &   50.74 &   -5 &   46 &    1.8 &    25.87 &  1225 &     211 &    -1 \\
   2338 &   -247 &      5 &   12 &   48 &   50.78 &   -5 &   46 &   12.9 &    27.39 &  1273 &     115 &    -1 \\
   2339 &   -232 &     20 &   12 &   48 &   49.44 &   -5 &   46 &    6.1 &    27.08 &  1285 &      -1 &    -1 \\
   2340 &   -272 &    161 &   12 &   48 &   47.88 &   -5 &   43 &   41.5 &    27.30 &  1239 &     357 &    -1 \\
   2341 &   -272 &     39 &   12 &   48 &   51.31 &   -5 &   45 &   32.3 &    27.70 &  1267 &      -1 &    -1 \\
   2342 &   -277 &     35 &   12 &   48 &   51.73 &   -5 &   45 &   33.5 &    27.74 &  1158 &      -1 &    -1 \\
   2343 &   -283 &     46 &   12 &   48 &   51.76 &   -5 &   45 &   21.5 &    -1.   &  1267 &     210 &    -1 \\
   2344 &   -290 &    111 &   12 &   48 &   50.37 &   -5 &   44 &   19.5 &    26.97 &  1110 &     336 &    -1 \\
   2345 &   -294 &    107 &   12 &   48 &   50.75 &   -5 &   44 &   21.5 &    26.55 &  1101 &     335 &    -1 \\
   2346 &   -338 &    143 &   12 &   48 &   52.37 &   -5 &   43 &   29.6 &    27.21 &  1120 &      -1 &    -1 \\
   2347 &   -327 &     19 &   12 &   48 &   55.21 &   -5 &   45 &   26.9 &    26.05 &  1425 &      -1 &    -1 \\
   2348 &   -361 &     81 &   12 &   48 &   55.54 &   -5 &   44 &   16.6 &    26.62 &  1360 &      -1 &    -1 \\
   2349 &   -373 &     96 &   12 &   48 &   55.84 &   -5 &   43 &   57.7 &    26.93 &  1238 &      -1 &    -1 \\
   2350 &   -403 &    101 &   12 &   48 &   57.51 &   -5 &   43 &   40.8 &    26.49 &  1105 &      -1 &    -1 \\
   2351 &   -409 &     95 &   12 &   48 &   58.08 &   -5 &   43 &   43.5 &    26.77 &  1229 &      -1 &    -1 \\
   2352 &    -83 &     17 &   12 &   48 &   40.48 &   -5 &   47 &   11.6 &    27.39 &  1286 &     224 &    -1 \\
   2353 &    -72 &     19 &   12 &   48 &   39.74 &   -5 &   47 &   14.3 &    27.61 &  1165 &     227 &    -1 \\

\enddata 
\tablenotetext{a}{The $x$, G and $y$, G pixel coordinates are given 
in arcsec and have their origin at the center of light of NGC 4697. 
The $x$, G coordinate is defined along the major axis of NGC 4697. 
The units of Right Ascension are hours, minutes and seconds; the 
units of Declination are degrees, arc minutes, and arc seconds.
In col. (6) a value of $-1$ for $m(5007)$ indicates that no measurement 
was attempted, because of CCD defects or a complex background. 
Heliocentric radial velocities are given in km s$^{-1}$. 
In the last two columns, a value of $-1$ indicates that the object either 
does not belong or was not measured in the corresponding FORS field.}
\end{deluxetable}

\clearpage

\begin{figure}
\epsscale{1.0}
\plotone{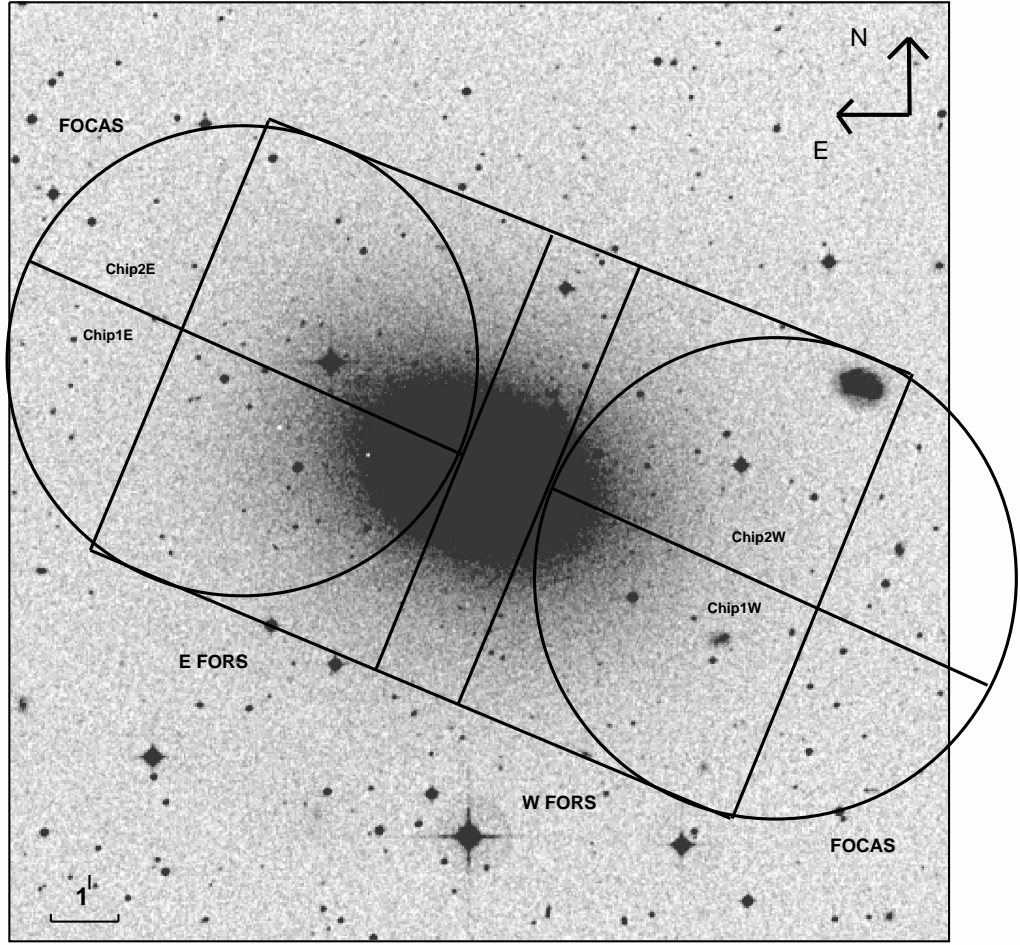}
\caption{
FOCAS (circular) and FORS (rectangular) fields
observed in NGC 4697. Paper I was based on observations of the FORS 
fields. The line splitting each circular FOCAS field indicates the 
position of the narrow gap (5 arcsec) separating chip 1 from chip 2.
}
\end{figure}

\begin{figure}
\epsscale{0.3}
\plotone{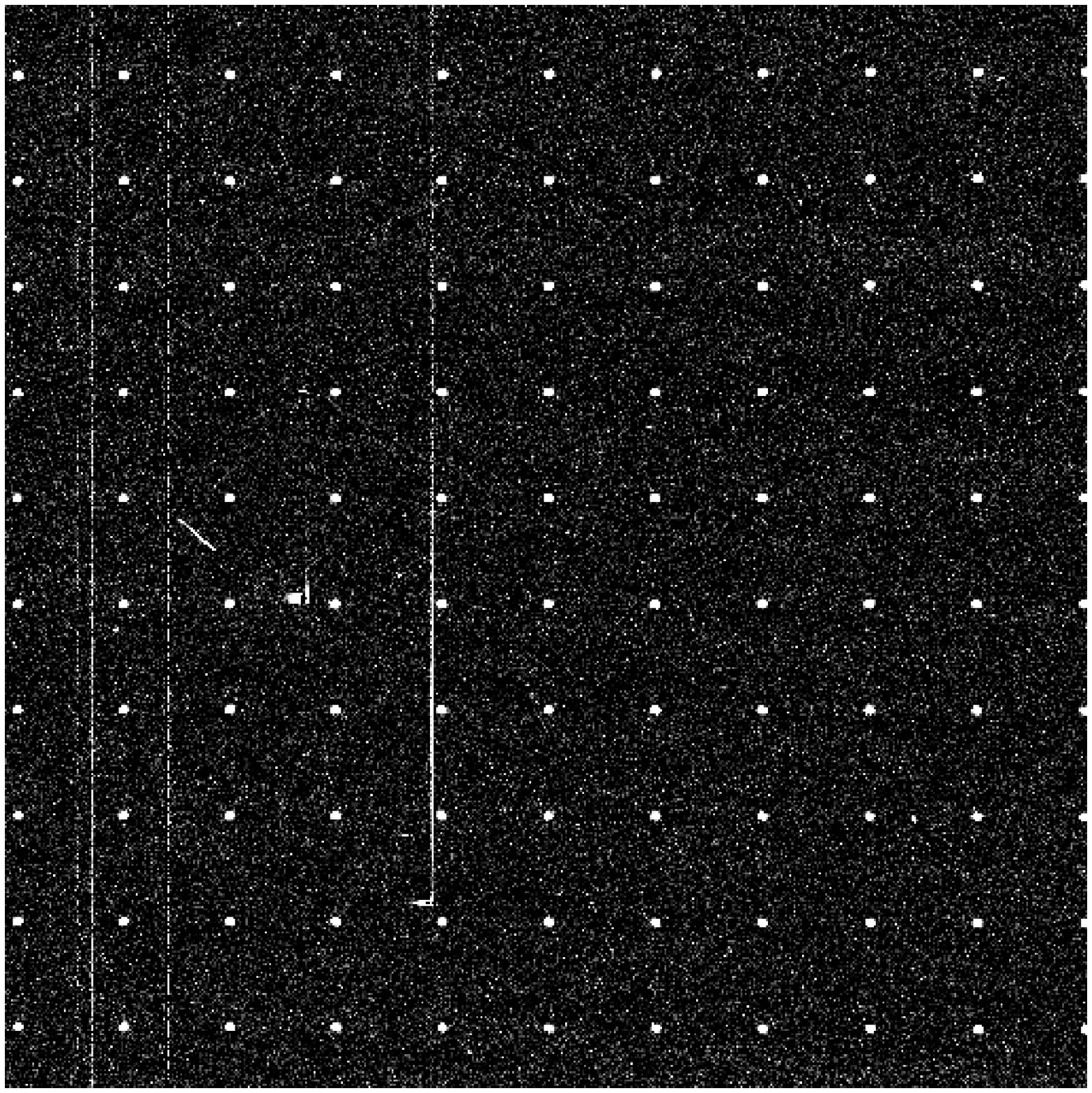}
\plotone{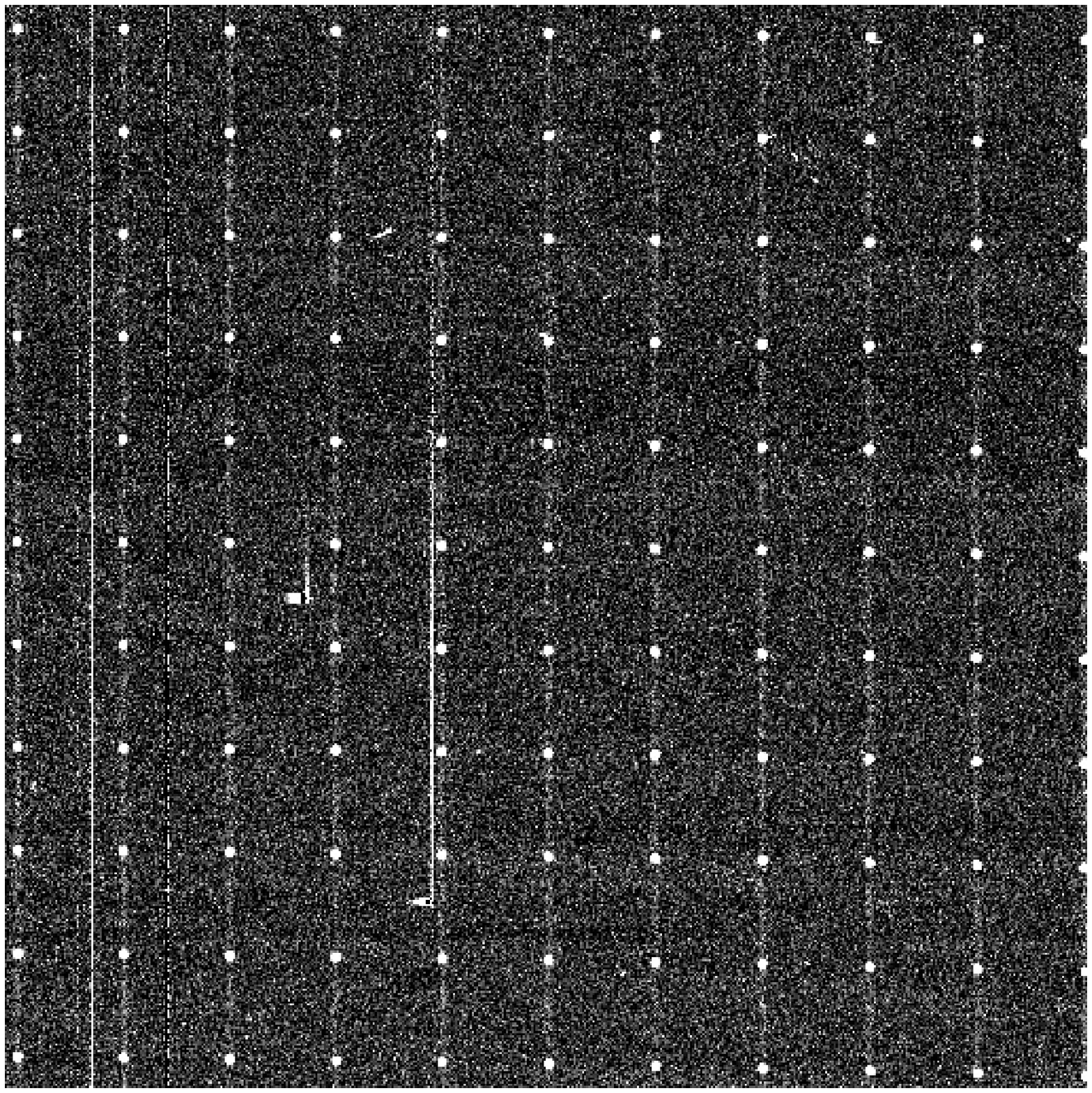}
\plotone{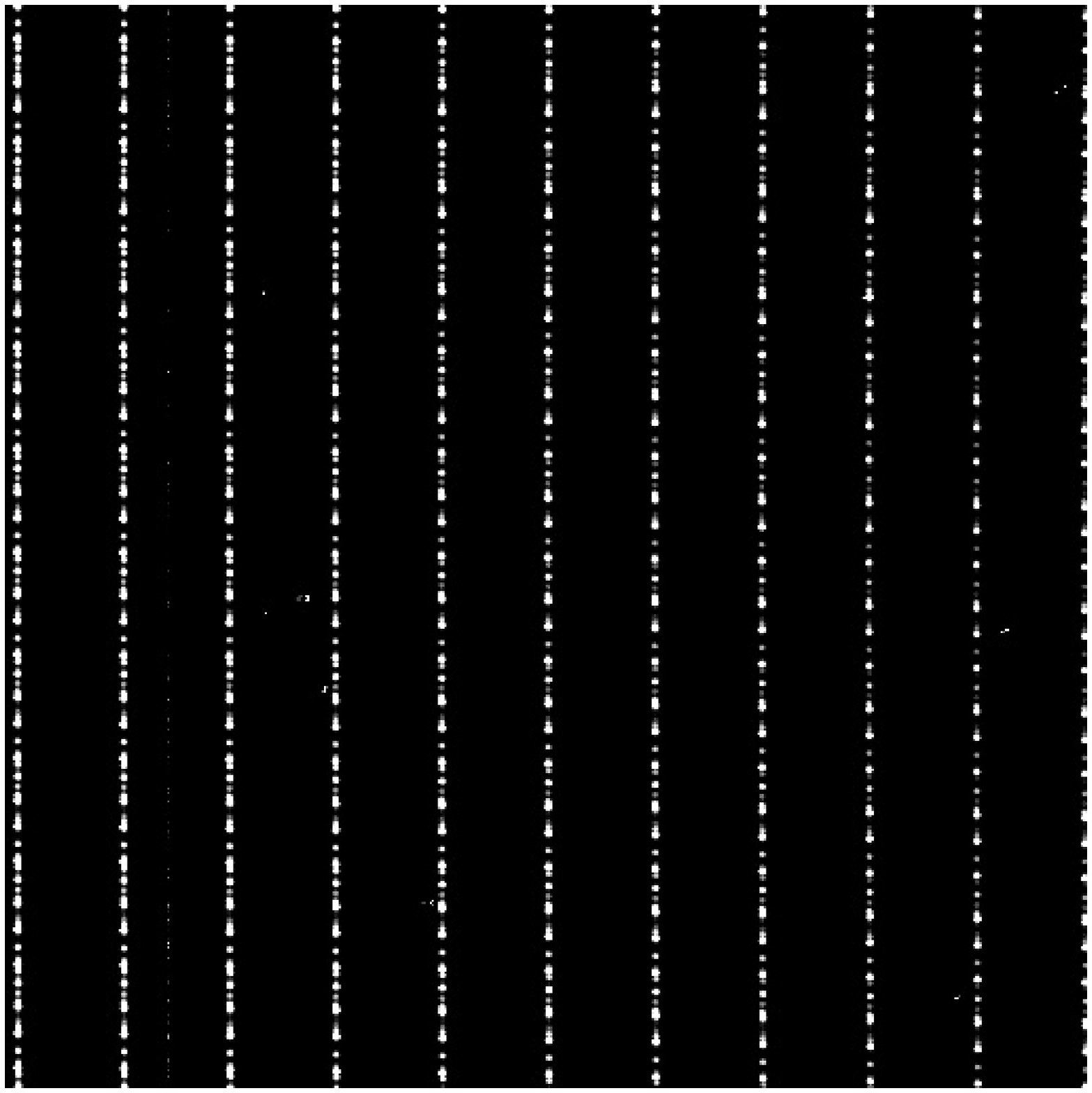}
\caption{
Calibration and quality control images obtained with FOCAS. From left to 
right, we show (u2) part of the engineering mask illuminated by NGC 7293, 
inserting the on-band filter; (d2) the same, adding the grism; what we 
see is [O III] $\lambda$5007 at each position (the only emission line 
passed by the on-band filter), plus a very faint nebular continuum; and 
(d1) the result of illuminating the mask with the Th-Ar comparison lamp, 
inserting filter N502 and grism. The direction of dispersion is vertical.
}
\end{figure}

\begin{figure}
\epsscale{1.0}
\plottwo{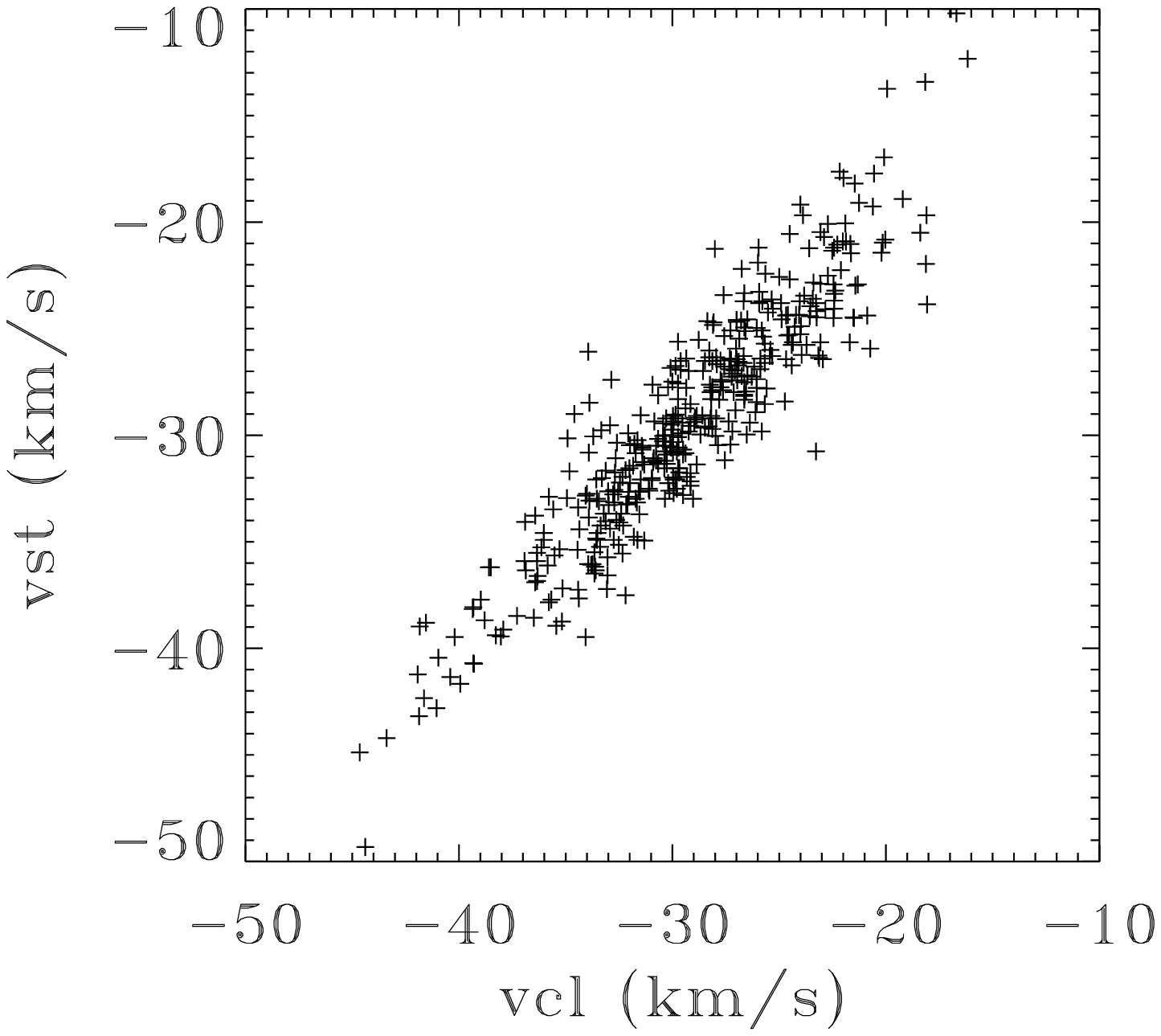}{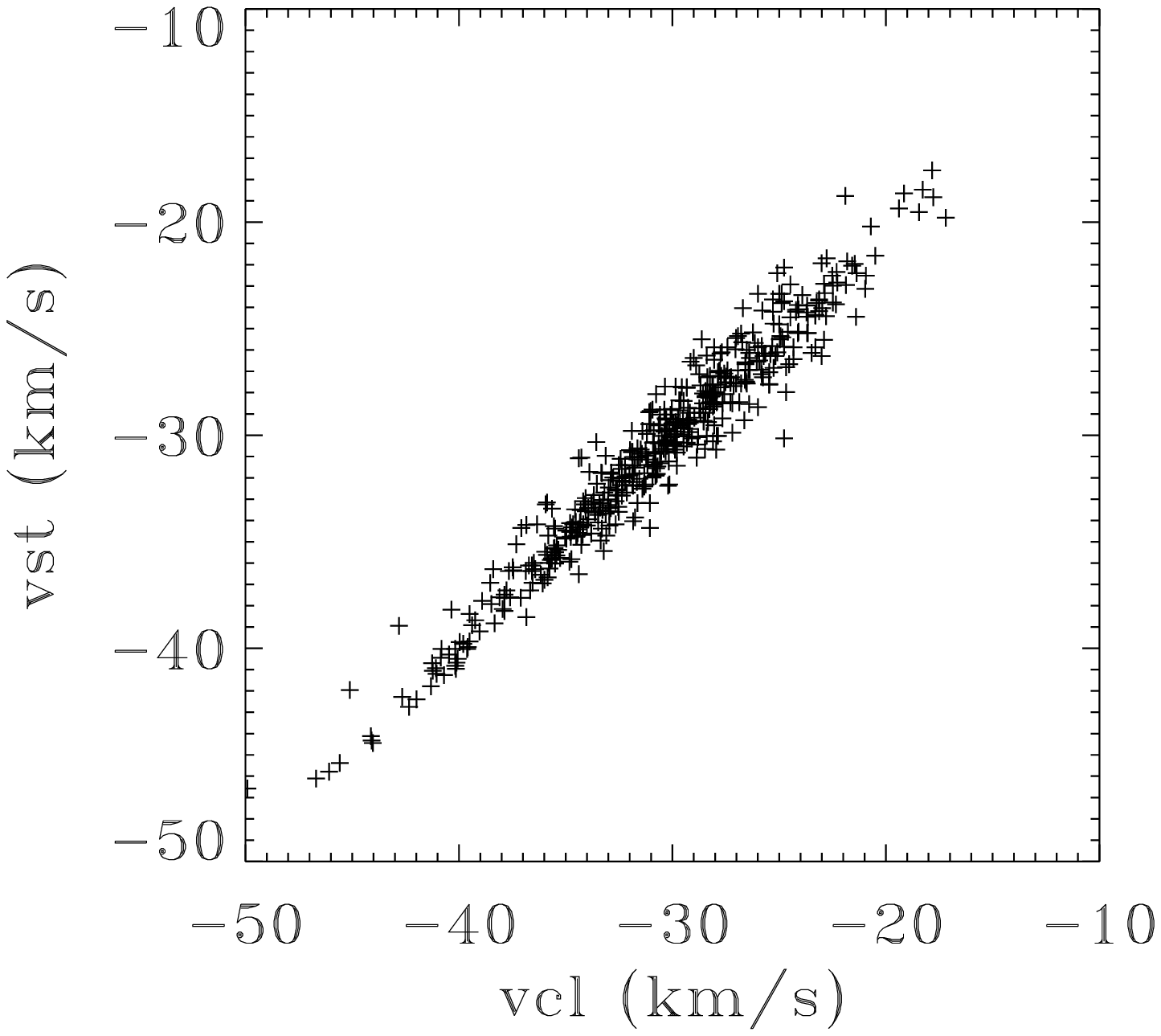}
\caption{
Comparison of slitless vs. classical FOCAS radial velocities of NGC 7293.
The left and right figures correspond to chips 1 and 2, respectively.
The PN's systemic radial velocity is $-$27 km s$^{-1}$, according to 
Meaburn et al. (2005).
}
\end{figure}

\begin{figure}
\epsscale{1.0}
\plotone{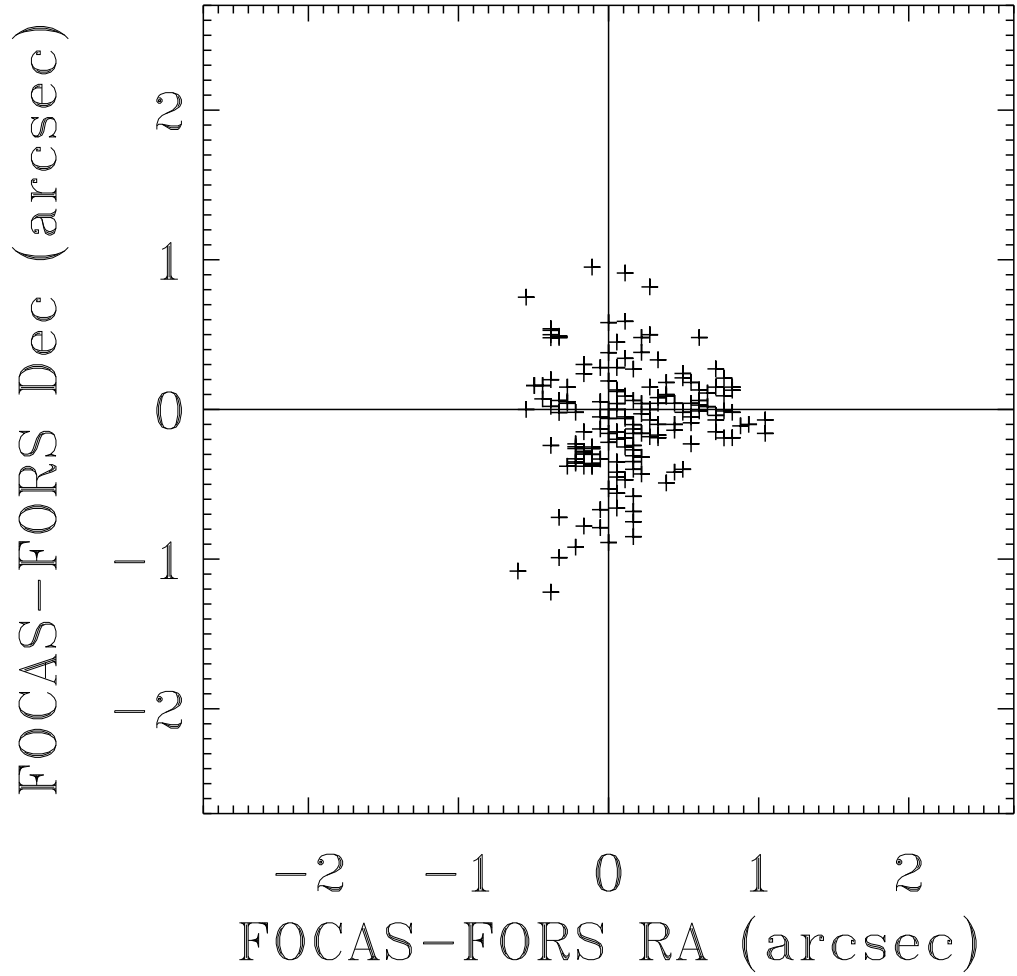}
\caption{
Comparison of equatorial coordinates for 162 PNs observed with both 
FORS and FOCAS.
}
\end{figure}

\begin{figure}
\epsscale{1.0}
\plotone{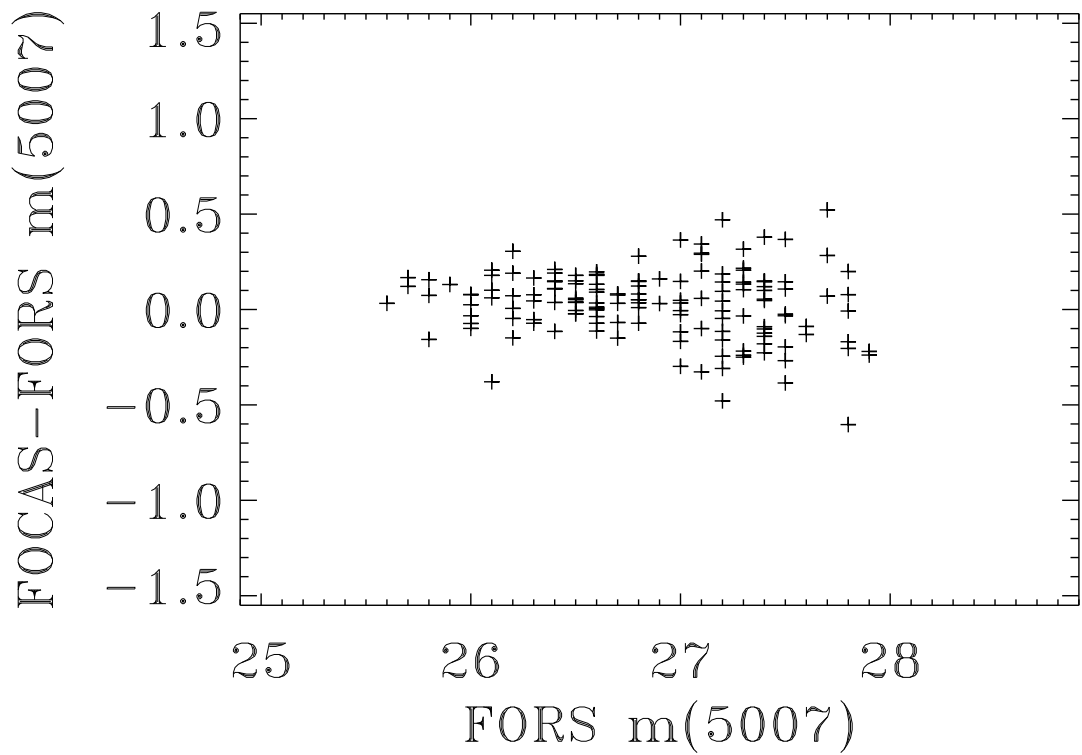}
\caption{
Differences (FOCAS $-$ FORS) of Jacoby magnitudes $m$(5007) plotted 
as a function of the FORS $m$(5007).
}
\end{figure}

\begin{figure}
\epsscale{0.3}
\plotone{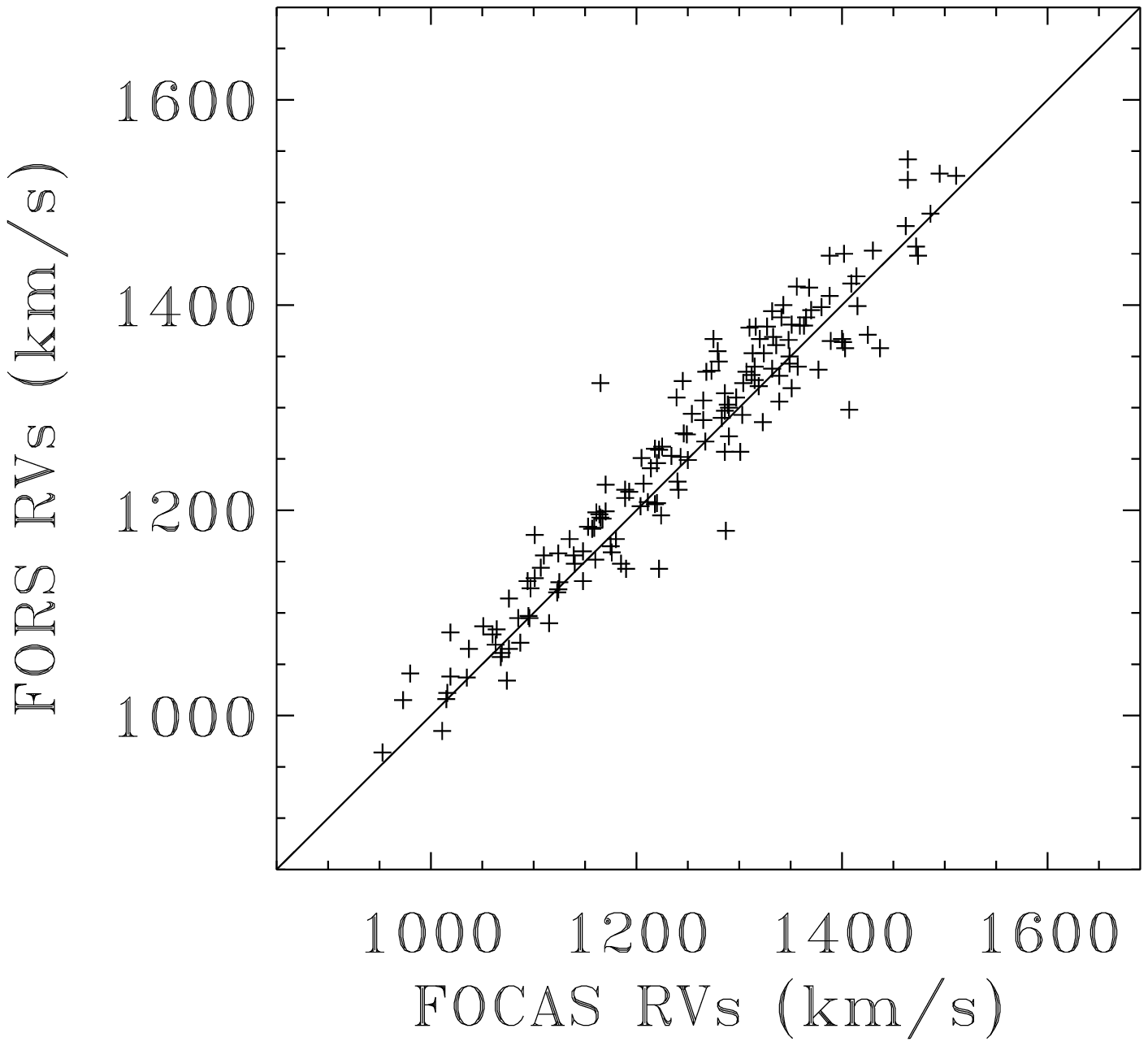}
\plotone{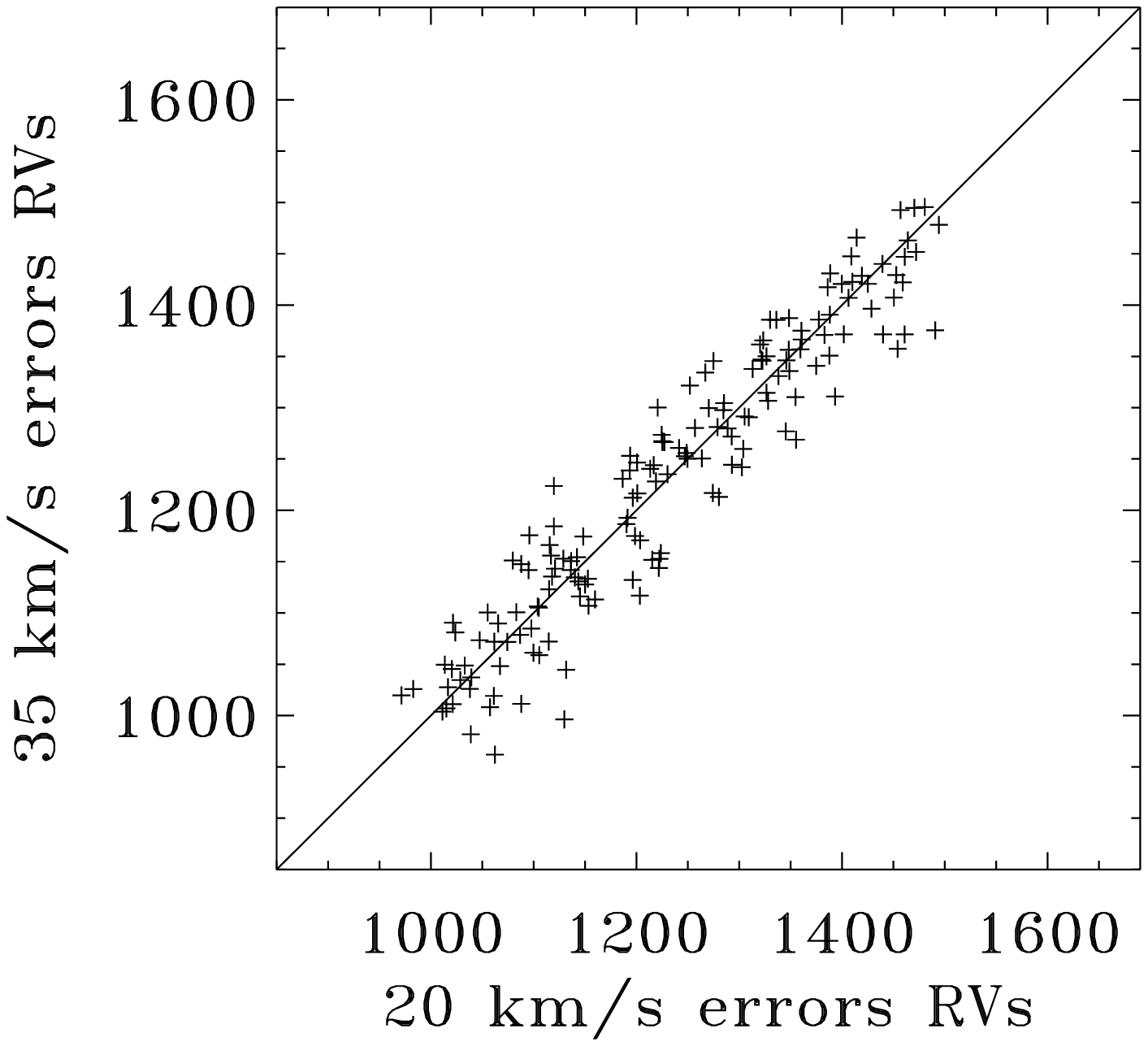}
\plotone{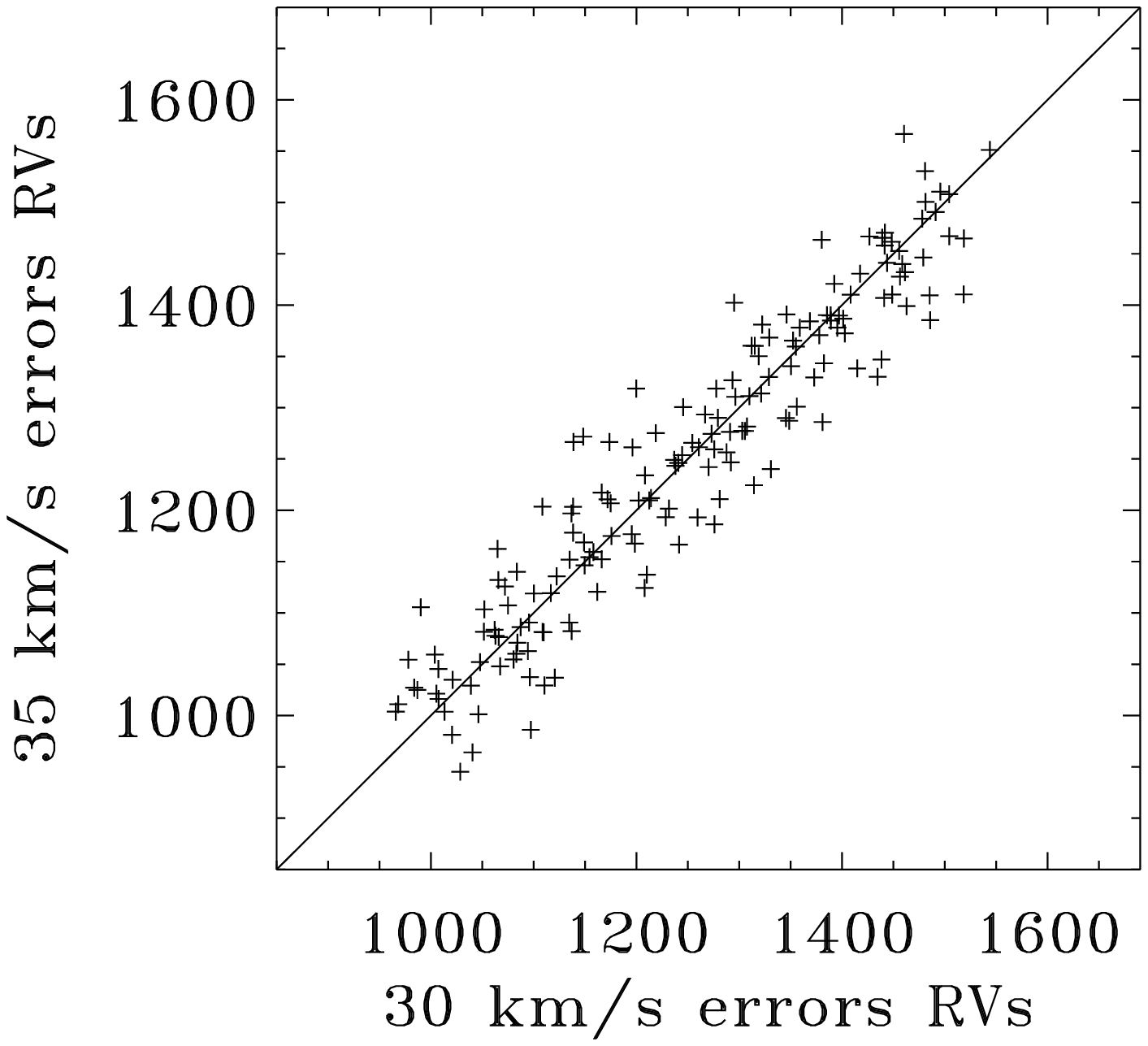}
\caption{
Left: comparison of heliocentric radial velocities for 158 PNs observed 
with both FORS and FOCAS. Center: random number simulations assuming
errors of 35 km s$^{-1}$ for FORS, and 20 km s$^{-1}$ for FOCAS. Right:
random number simulations assuming 30 km s$^{-1}$ for FOCAS. We confirm
that the FOCAS errors are not larger than 20 km s$^{-1}$.
}
\end{figure}

\begin{figure}
\epsscale{1.0}
\plotone{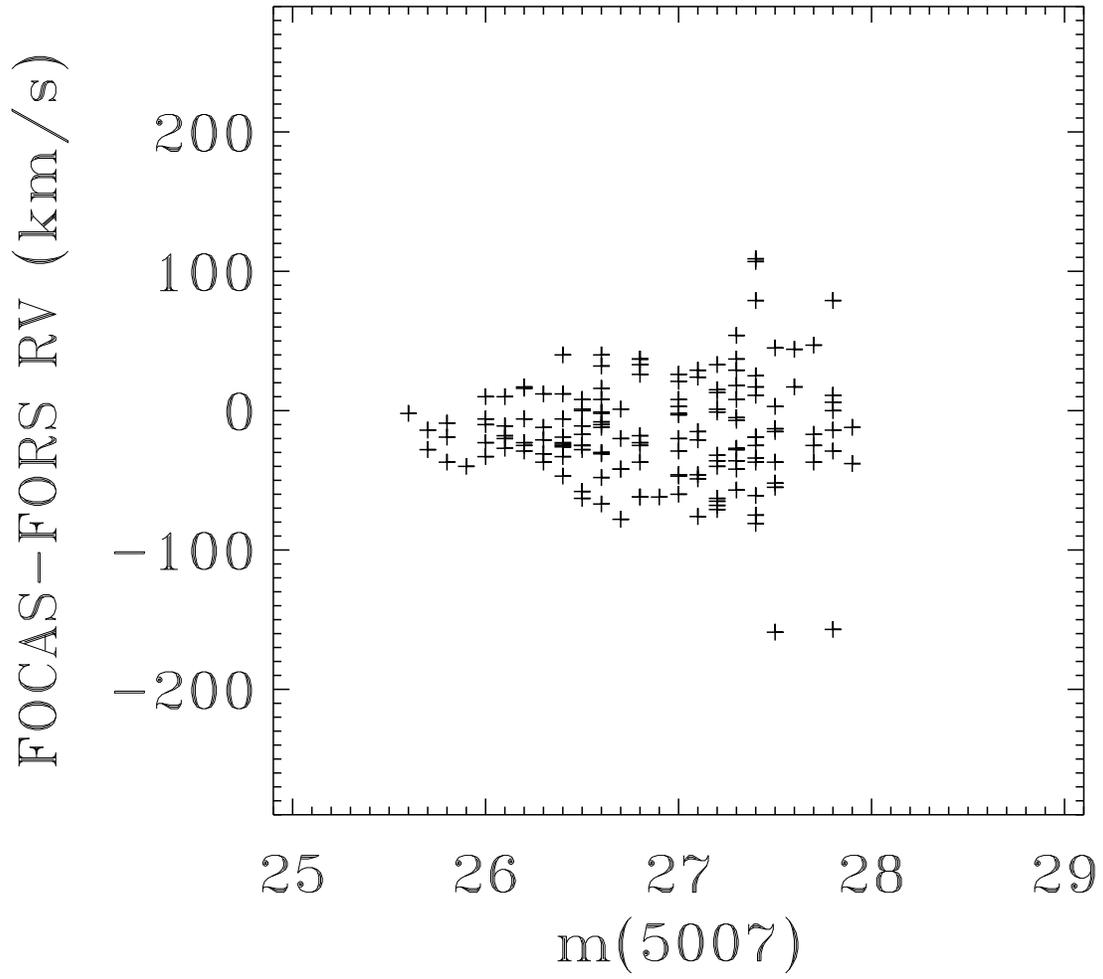}
\caption{
Velocity differences (FOCAS $-$ FORS) plotted as a function of $m$(5007).
}
\end{figure}

\begin{figure}
\epsscale{1.0}
\plottwo{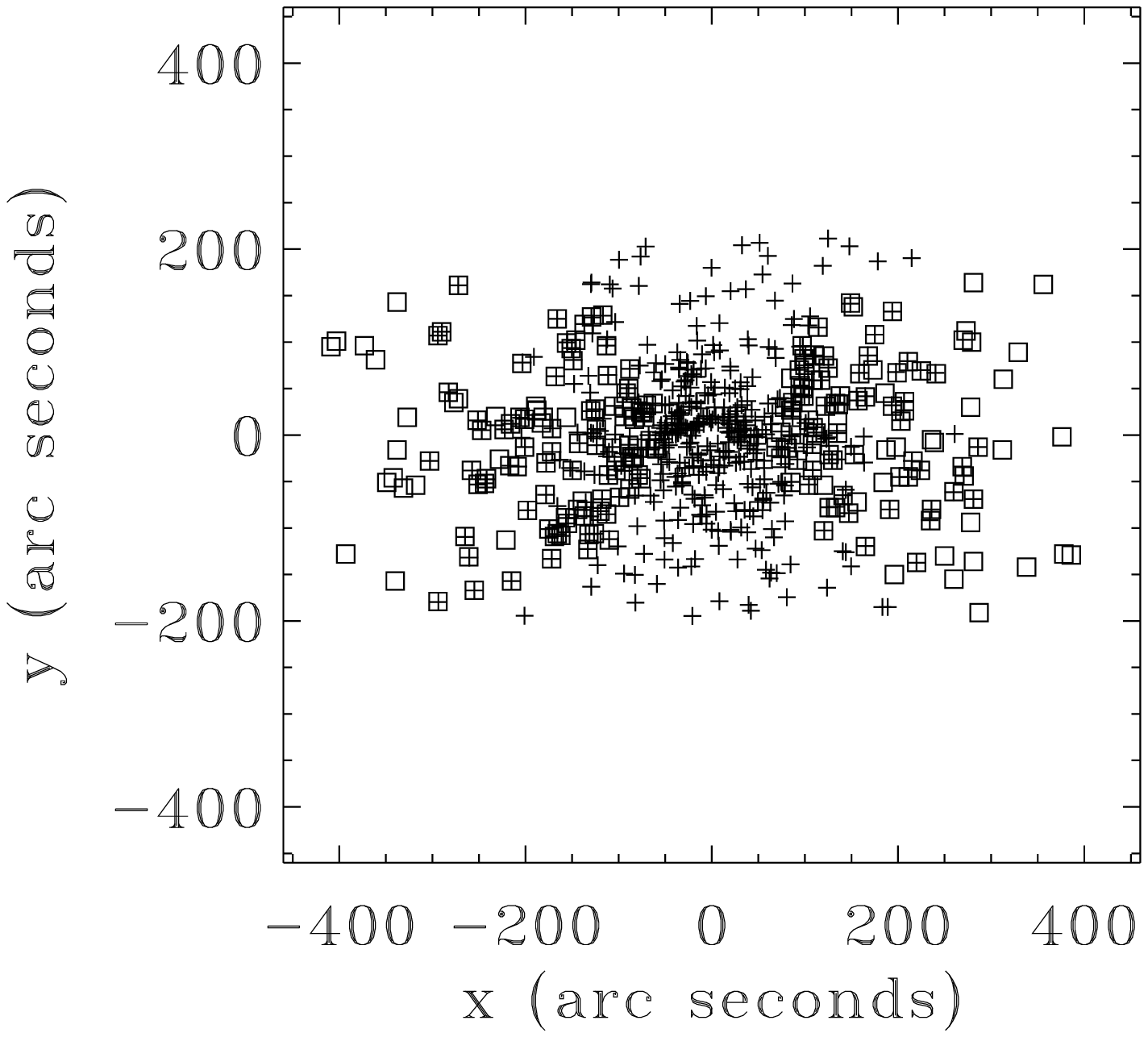}{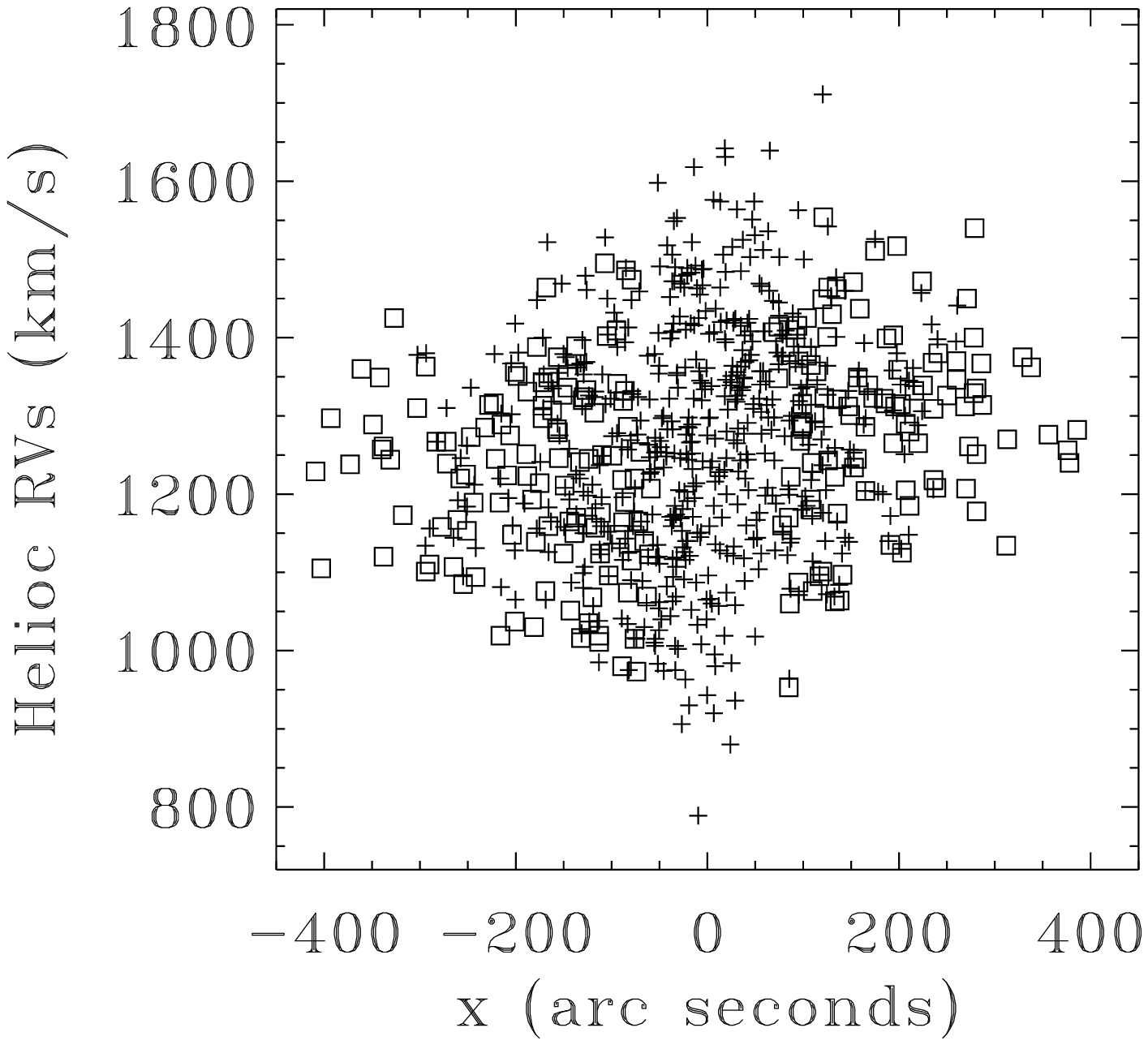}
\caption{
FORS (plus signs) and FOCAS (squares) PN detections.
Left: positions of PNs relative to the center of light of NGC 4697, in 
arcsec. Right: radial velocities as a function of the $x$ coordinates. 
}
\end{figure}

\begin{figure}
\epsscale{1.0}
\plotone{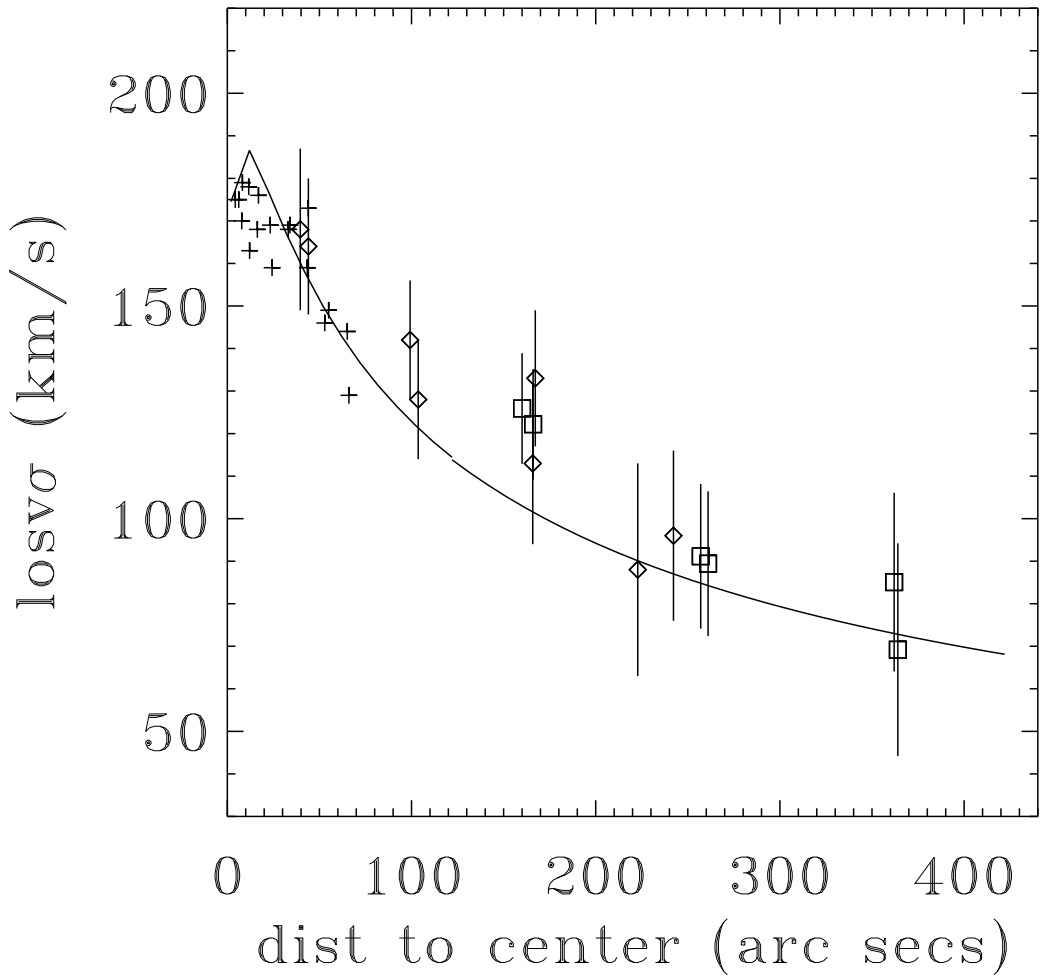}
\caption{
Line-of-sight velocity dispersion plotted as a function of average 
angular distance from the center of NGC 4697. Plus signs are 
measurements by Binney et al. (1990) on long-slit, integrated-light 
spectra along the major axis. Diamonds are PN results from Paper I.
Squares are PN results from FOCAS radial velocities.
The solid line is the Hernquist (1990) analytical model with
a constant $M/L$ ratio, adopting $R_{\rm e} =$ 66 arcsec,
and a total mass of 1.5$\times 10^{11} M_\odot$. This is equivalent 
to $(M/L)_B = 9$.
}
\end{figure}

\begin{figure}
\epsscale{1.0}
\plotone{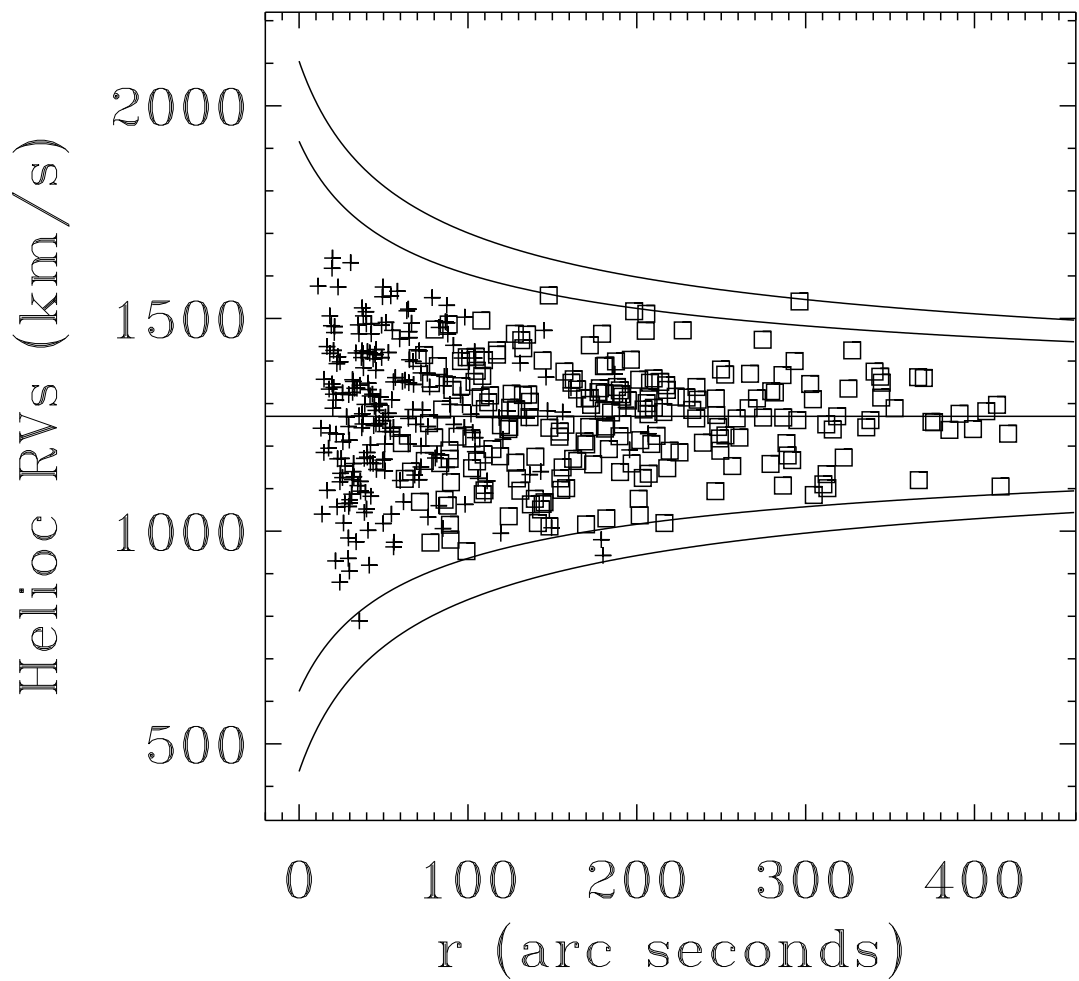}
\caption{
Individual FOCAS (squares) and FORS (plus signs) PN radial velocities 
plotted as a function of angular distance from the center of NGC 4697.
We have only used FORS velocities of PNs from regions near the center, 
where we do not have FOCAS velocities.
The solid lines are escape velocities for Hernquist models with 
$(M/L)_B =$ 9 (outer) and 5 (inner).
}
\end{figure}

\begin{figure}
\epsscale{0.3}
\plotone{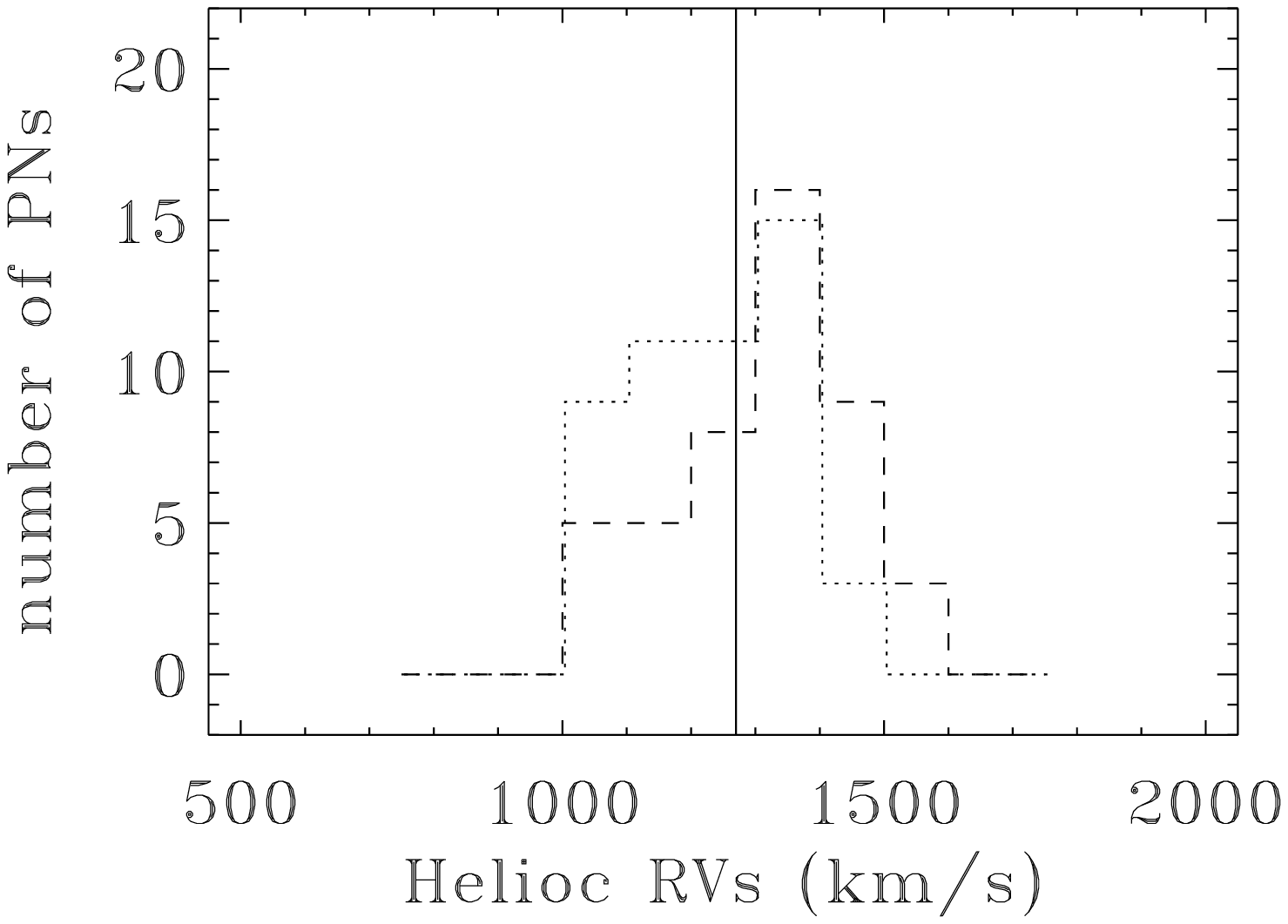}
\plotone{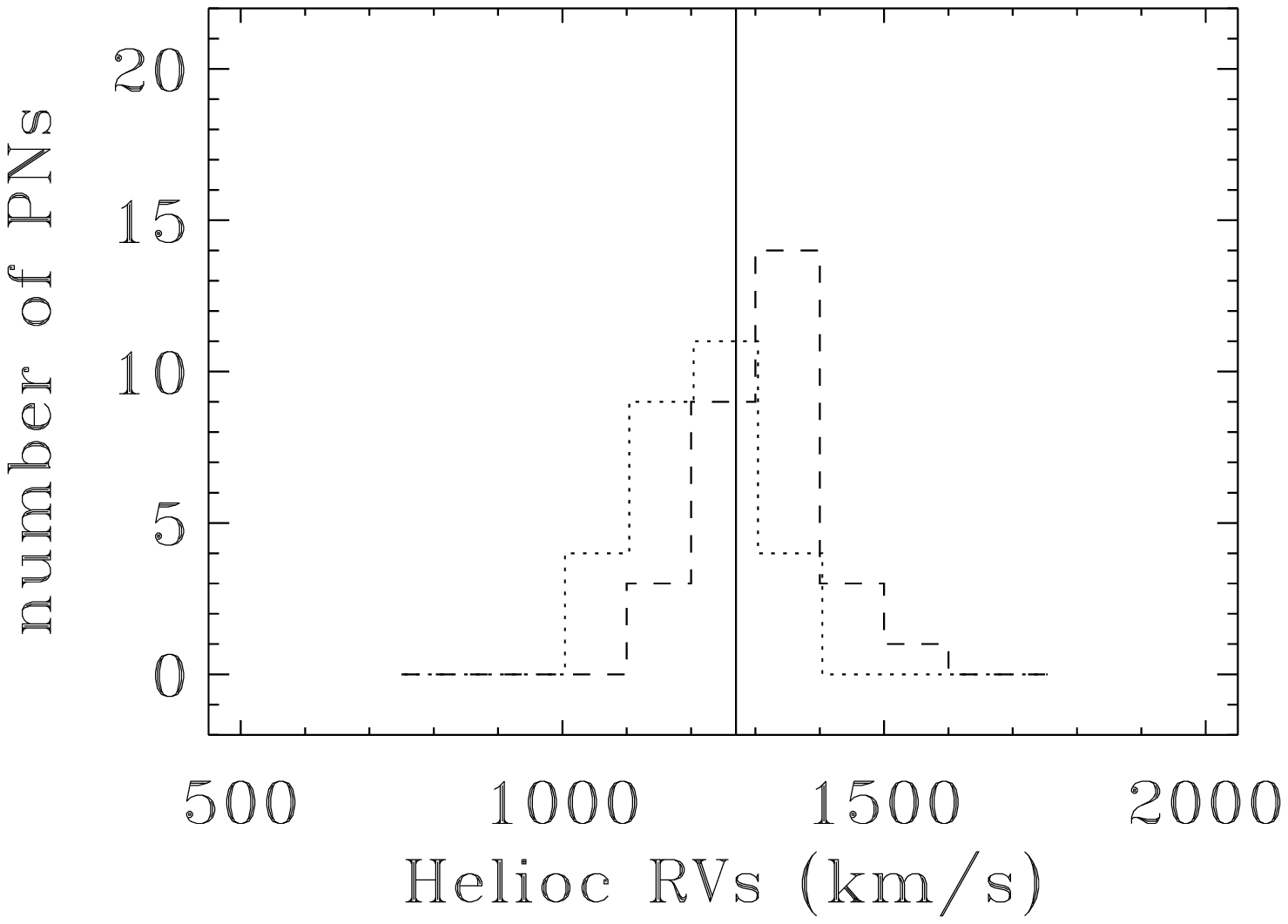}
\plotone{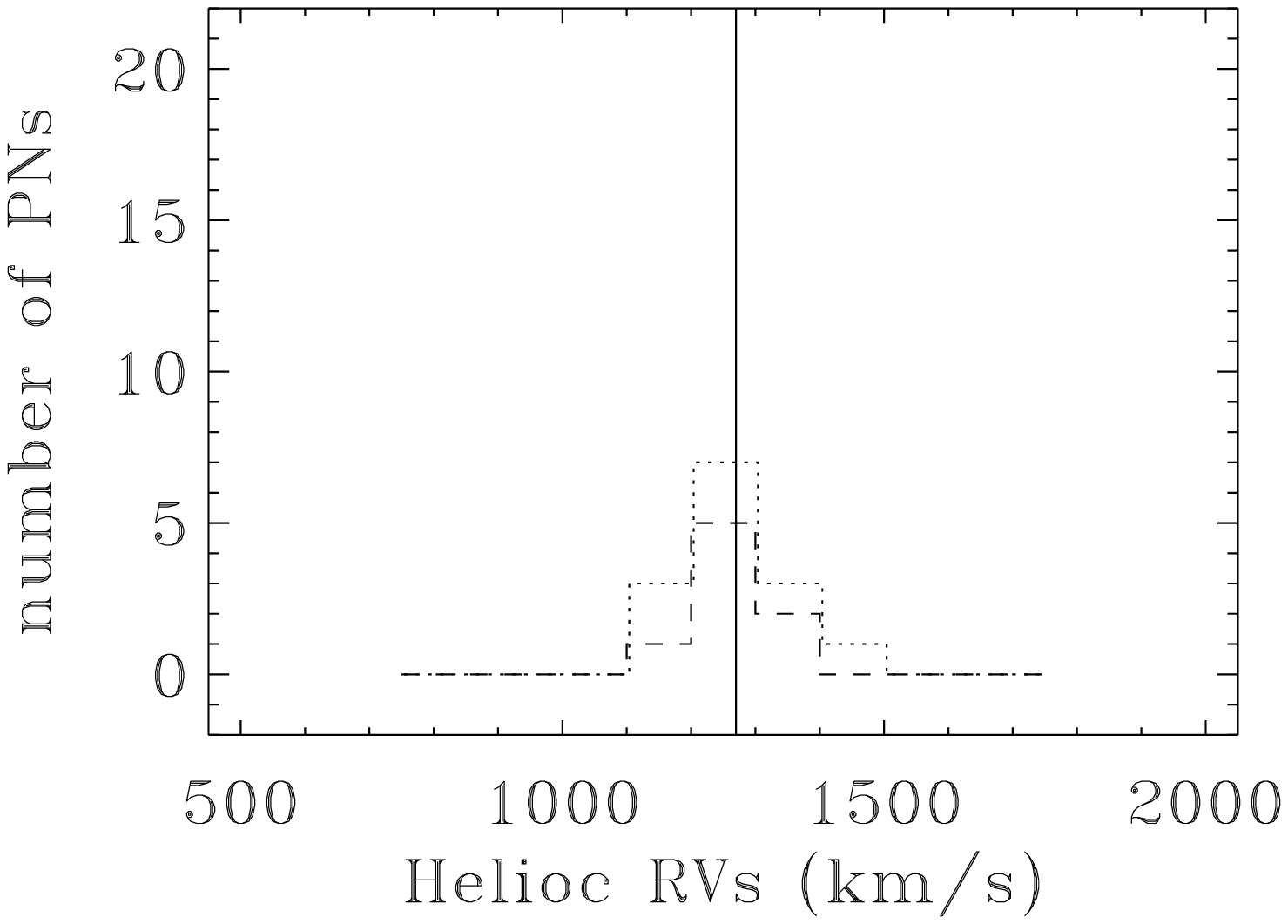}
\caption{
Left panel:
radial velocity (FOCAS) histograms for PNs with $-200 < x < -100$ 
(dotted line) and $100 < x < 200$ (dashed line). The vertical line 
indicates the systemic velocity, 1270 km s$^{-1}$, taken from Paper I.
There is an excess of positive relative velocities for negative $x$, but
still the rotational signal is clear: at positive $x$ the PNs are 
preferentially moving away from us (in the same sense as the stars),
while the opposite happens at negative $x$. 
Central panel:
radial velocity (FOCAS) histograms for PNs with $-300 < x < -200$ 
(dotted line) and $200 < x < 300$ (dashed line). The vertical line 
has always the same meaning. As we move outward, there is still a clear 
rotational signal (compare with the next panel).
Right panel:
radial velocity (FOCAS) histograms for PNs with $x < -300$ (dotted line)
and $300 < x$ (dashed line). No rotation is apparent here.
}
\end{figure}

\begin{figure}
\epsscale{1.0}
\plotone{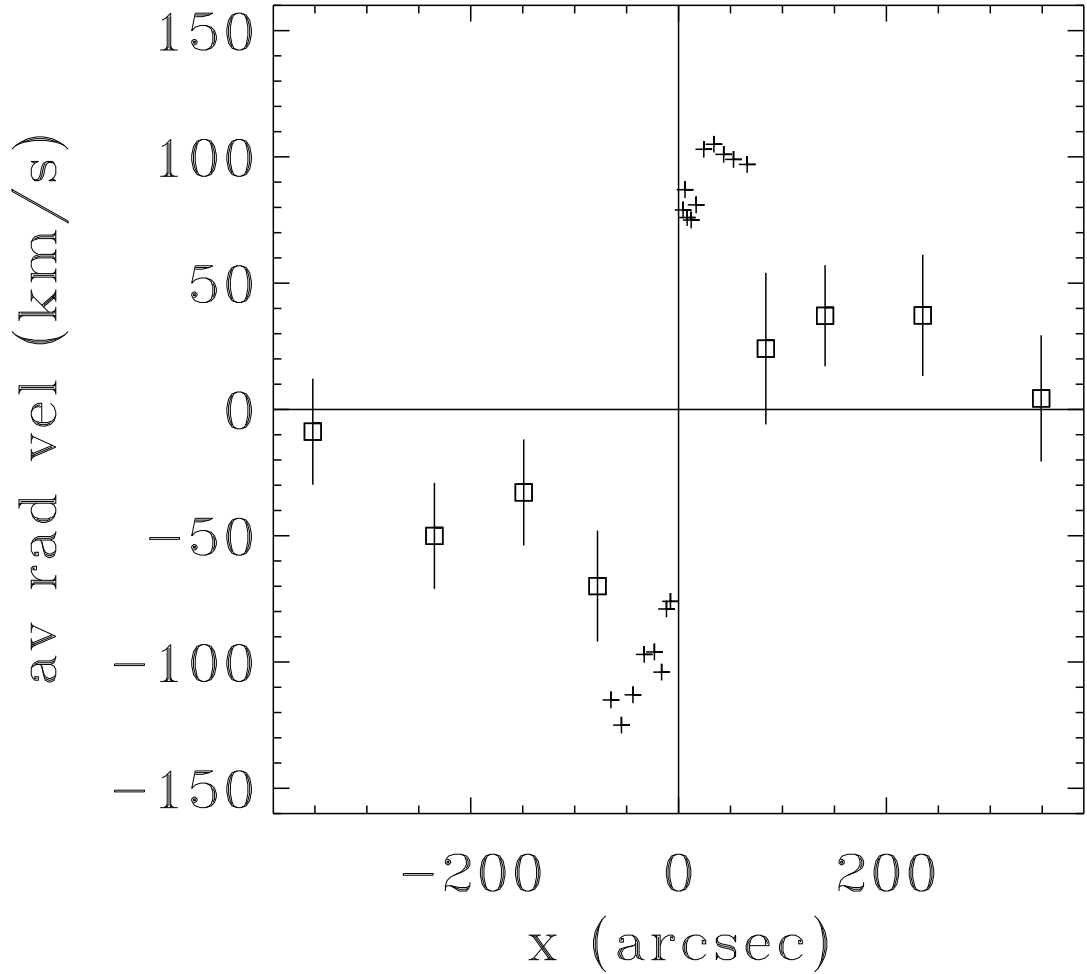}
\caption{
Average FOCAS PN radial velocity (squares) plotted as a function 
of average $x$ coordinate for eight PN groups, described in the text. 
Plus signs are velocities measured on long-slit, integrated-light 
spectra along the major-axis by Binney et al. (1990). The origin of 
velocities corresponds to the systemic velocity of NGC 4697, 
1270 km s$^{-1}$. 
The effective radius of NGC 4697 is 66 arcsec.
}
\end{figure}

\end{document}